\documentclass[english,twocolumn,transaction]{IEEEtran}
\usepackage[T1]{fontenc}
\usepackage{float}
\usepackage{amsmath}
\usepackage{amsthm}
\usepackage{amssymb}
\usepackage{stackrel}
\usepackage{graphicx}
\usepackage{color}

\makeatletter

\floatstyle{ruled}
\newfloat{algorithm}{tbp}{loa}
\providecommand{\algorithmname}{Algorithm}
\floatname{algorithm}{\protect\algorithmname}

\theoremstyle{plain}
\newtheorem{thm}{\protect\theoremname}
\theoremstyle{plain}
\newtheorem{prop}[thm]{\protect\propositionname}
\newtheorem{lemma}{Lemma}
\IEEEoverridecommandlockouts
\usepackage{ifpdf}
\usepackage{cite}
\ifCLASSOPTIONcompsoc
\usepackage[caption=false,font=normalsize,labelfont=sf,textfont=sf]{subfig}
\else
\usepackage[caption=false,font=footnotesize]{subfig}
\fi
\usepackage{amsmath}
\usepackage{flushend}
\@ifundefined{showcaptionsetup}{}{%
\PassOptionsToPackage{caption=false}{subfig}}
\usepackage{subfig}

\makeatother

\usepackage{babel}
\providecommand{\propositionname}{Proposition}
\providecommand{\theoremname}{Theorem}

\begin{document}
\renewcommand\figurename{Fig.}
\title{Secure Beamforming Design in Relay-Assisted Internet of Things}

\author{~Pingmu~Huang,~Yunqin~Hao,~Tiejun~Lv,~\IEEEmembership{Senior Member,~IEEE},~Jintao Xing,~Jie~Yang,\\~and P. Takis Mathiopoulos, \emph{Senior Member, IEEE}
\thanks{The financial support of the National Natural Science Foundation of China (NSFC) (Grant No. 61671072) is gratefully acknowledged. (\emph{Corresponding author: Tiejun Lv.})

P. Huang, Y. Hao, T. Lv, J. Xing and J. Yang are with the School of Information and Communication Engineering, Beijing University of Posts and Telecommunications (BUPT), Beijing 100876, China (e-mail: \{pmhuang, haoyunqin, lvtiejun, jintaoxing, janeyang\}@bupt.edu.cn).

P. T. Mathiopoulos is with the Department of Informatics and Telecommunications,
National and Kapodistrian University of Athens, Athens 157 84, Greece
(e-mail: mathio@di.uoa.gr).
}}
\maketitle
\begin{abstract}

A secure downlink transmission system which is exposed to multiple eavesdroppers and is appropriate for Internet of Things (IoT) applications is considered. A worst case scenario is assumed, in the sense that, in order to enhance their interception ability all eavesdroppers are located close to each other, near the controller and collude to form joint receive beamforming. For such a system, a novel cooperative non-orthogonal multiple access (NOMA) secure transmission scheme for which an IoT device with a stronger channel
condition acts as an energy harvesting relay in order to assist a second IoT device operating under weaker channel conditions, is proposed and its performance is analyzed and evaluated. A secrecy sum rate (SSR) maximization problem is formulated and solved under three constraints: i) Transmit power; ii) Successive interference cancellation; iii)  Quality of Service. By considering both passive and active eavesdroppers scenarios, two optimization schemes are proposed to improve the overall system SSR.  On the one hand, for the passive eavesdropper scenario, an artificial noise-aided secure beamforming scheme is proposed. Since this optimization problem is non-convex, instead of using traditional but highly complex, brute-force two-dimensional search, it is conveniently transformed into a convex one by using an epigraph reformulation. On the other hand, for the active multi-antennas eavesdroppers' scenario, the orthogonal-projection-based beamforming scheme is considered, and by employing the successive convex approximation method, a suboptimal solution is proposed. Furthermore, since for single antenna transmission the orthogonal-projection-based scheme may not be applicable a simple power control scheme is proposed.
Various performance evaluation results obtained by means of computer simulations have verified that the proposed schemes outperform other benchmark schemes in terms of SSR performance.

\end{abstract}

\section{Introduction}

The Internet of Things (IoT) is rapidly evolving as a complex platform connecting a very large number of communication devices, e.g. sensors, controllers and actuators \cite{IoT}. However, achieving the required ubiquitous connectivity required for such IoT based communication  systems becomes a vital and challenging task mainly because of  the constraint of scarce bandwidth. In this context, nonorthogonal multiple access (NOMA) is advocated as a promising technique to support pervasive connectivity, so that the spectral efficiency of IoT systems can be significantly enhanced \cite{Secrecy_sum_rate,Min_max_NOMA,Ergodic_capacity,Magazine,SCA}.
The key feature of NOMA is to implement the multiple access (MA) in the power domain while the time/frequency/code resources can simultaneously be shared by all users. Moreover, as compared to orthogonal MA (OMA), NOMA can achieve a better balance between sum rate and user fairness
\cite{fairness}.
At the same time, as the wireless broadcast nature of radio propagation makes IoT communications susceptible to eavesdropping attacks, their security aspects need to be very carefully considered when designing such systems. In the past,
conventional encryption techniques have been applied to prevent eavesdroppers
from recovering the secret messages \cite{traditional1,traditional2}. However, such encryption-based
techniques have inherent difficulties and vulnerabilities dealing with secret
key management \cite{IoT}. Fortunately, physical layer
security (PLS) has shown a great deal of potential to more accurately distinguish the signals belonging to the legitimate receiver or the eavesdropper \cite{enlarge,cooperative_1,cooperative_2}. Unlike other encryption-based techniques, PLS techniques take advantage of the physical characteristics of wireless medium to ensure information-theoretic security independent of the eavesdropper's computing capabilities \cite{cooperative_3,cooperative_5}.

\subsection*{A. Motivation and Contributions}
In the past, several PLS techniques based on the OMA secure transmission protocol, including  multi-antennas
techniques  \cite{Eve_secrecyrate, ECSI}, artificial noise (AN) \cite{AN,AN_1,AN_2,AN_3, my_paper, ZF_constraint, AN_number} and  cooperative jamming
(CJ) \cite{CJ, Eve_secrecysum, cooperative_jamming} have been proposed. Aiming to further improve the secrecy of the fifth generation (5G)
systems, Tian \textit{et al}. \cite{Secrecy_sum_rate} combined NOMA
with multi-antennas techniques to optimize the secrecy sum rate (SSR)
of 5G wireless system for applications where the channel state information (CSI) of
eavesdroppers is available. In another approach, Hu \textit{et al}. \cite{IoT} have combined the AN assisted multi-antennas transmission technique with CJ in order to reduce the effects of the passive non-colluding eavesdroppers for the IoT downlink.
However, since many IoT device transmitters have limited radio frequency (RF) power, allocating some power for AN or jamming signal may not be  appropriate since this will restrict  the coverage range of secure transmissions.

In order to fill this gap, in this paper a novel relay-assisted secure transmission scheme is proposed designed for energy-constrained IoT based communication systems, where one controller emits secret messages
to two classes of devices operating in the presence of multiple colluding eavesdroppers.
These IoT devices have diversified quality of service (QoS) requirements, and the channel conditions among them are quite different in the sense that the channel operating conditions for $D_{1}$ are much better then the channel conditions for $D_{2}$. In order to guarantee
the confidentiality of data transmission and the QoS requirement of
$D_{2}$, $D_{1}$ acts as an  energy
harvesting (EH) relay to help
$D_{2}$. Assuming that the locations of all eavesdroppers are close to the controller, this worst-case positioning scheme will increase the probability of interception of the information signal. For this energy-constraint IoT system, a novel secure cooperative transmission scheme is proposed and its performance is analyzed and evaluated. Assuming that the CSI of the eavesdroppers is available, two operational scenarios are considered, namely the passive eavesdroppers and active eavesdroppers scenarios, for which two optimization. For maximizing the SSR of the IoT system under consideration under transmit power, QoS and successive interference cancellation (SIC) constraints. Within this framework, the primary contributions of this paper
are summarized as follows:
\begin{itemize}
\item A cooperative simultaneous wireless information and power transfer (SWIPT) secure transmission protocol which takes into account the diversified QoS requirements of different IoT users, is proposed. Unlike secure transmission designs without SWIPT\cite{IoT,Secrecy_sum_rate}, by employing the SWIPT-aided  transmission protocol, device 1 forwards the
information bearing signal and AN without introducing  additional energy consumption.
\item For the passive eavesdroppers scenario, an AN-aided secure beamforming
scheme which jointly optimizes the transmit beamforming vectors, power splitting (PS) ratio and  AN beamforming, is proposed. Since this optimization
problem is non-convex, it is conveniently transformed into a convex form through the use of a
epigraph reformulation (ER), instead of using the two-dimensional search
\cite{two-dimensional-search} that requires high computational complexity.
\item For the active  eavesdroppers scenario, an orthogonal-projection-based
secure beamforming design scheme which ensures the confidentiality
of data transmission is proposed. As this challenging optimization problem is also non-convex, the NP-hard problem is reformulated into a trackable convex problem.
Furthermore, for single antenna transmitters,
the orthogonal-projection-based beamforming design problem simplifies to a power allocation problem which should not be solved via the orthogonal projection based method.
Instead, a power control scheme is proposed through which via a transformation of the objective
function and constraints the non-convex problem can be efficiently
handled.
\end{itemize}

\subsection*{B. Related Previous Research}

There has been recently a lot of interest from the information-theoretic point of view, to exploit PLS techniques in order to improve the security of wireless communication systems \cite{my_paper,AN_number,CJ}. For example, considering multiple-input single-output channels (SISO), the cases of both direct transmission and CJ with a helper have been investigated under the assumption that CSI for the wiretap links is imperfect \cite{ZF_constraint}. For multiple-input multiple-output (MIMO) channels, \cite{Eve_secrecysum} proposed a CJ-aided SSR optimization scheme under the assumption that the power of the system is sufficient. Furthermore, \cite{AN_number,cooperative_jamming} have focused on studying the  confidential information transmission of relay systems with the PLS techniques-aided. The secure transmission schemes introduced in these two papers are based on the assumption that available power of the considered relay systems is sufficient. More recently in \cite{my_paper},  a robust secure beamforming design  for MIMO two-way relaying system based on the AN and physical layer network coding has been introduced. Various performance evaluation results presented in \cite{my_paper} have shown that such beamforming design will further improve the security aspects of the two-way relaying system.

It is underlined that the above mentioned papers (\cite{my_paper,AN_number,CJ,ZF_constraint,Eve_secrecysum,cooperative_jamming}) investigate various PLS issues encountered in different cooperative wireless communication systems  without considering possible energy-limited constraints. Furthermore, these papers have not
investigated the security transmission problem in connection with IoT type of wireless communication systems. On this topic, \cite{IoT,traditional2,cooperative_5},  different PLS-based secure transmission schemes have been presented for IoT systems. However these papers have not considered the energy-limited feature of IoT systems when designing the secure transmission beamforming.

To this end, it is necessary to solve  the security problem of the energy-constrained IoT systems with EH-assisted. In this paper SWIPT  is considered as an EH technique where the IoT transmitter forwards information signal and AN via the harvested energy only.
Our work is different from above works in the following aspects: i) \cite{IoT} investigates the secure transmission problem of IoT system, from an information-theoretic point of view, and only the CJ technique is used so that the coverage range of secure transmissions couldn't be guaranteed; ii) \cite{traditional2} considers a variety of eavesdropping scenarios when designing the secure transmission protocols, but colluding eavesdroppers cases are not explored; iii) \cite{cooperative_5} considers a secure transmission for IoT under eavesdropper collusion. Although the scheme proposed in \cite{cooperative_5} is more general than that the schemes presented in \cite{IoT} and \cite{traditional2}, it has not considered secure beamforming design nor the energy-constrained problem. Note in our research here we not only provide secure beamforming designs, but also the energy-constrained problem is considered by combining the PLS technique with SWIPT.

\subsubsection*{C. Notation and Organization}

The remainder of this paper is organized as follows: In Section II,
we describe the system model and problem formulation. In Section III,
AN-aided transmission beamforming design scheme is presented when eavesdroppers'
CSI is unavailable. In Section IV, by taking both the single-antenna
and multiple-antenna scenarios into account, two SSR optimization
schemes are proposed when the eavesdropper's CSI is available. The
performance evaluation results and discussion are provided in Section IV. The conclusions of the paper can be found in Section VI.

Boldface lowercase and uppercase letters denote vectors and matrices,
respectively. The Hermitian transpose, Frobenius norm,  and trace of
the matrix $\mathbf{A}$ are denoted as $\mathbf{A}^{H}$, $\left\Vert \mathbf{A}\right\Vert $, and $\mathrm{Tr}(\mathbf{A})$, respectively, whereas rank$\left(\mathbf{A}\right)$ and $|\mathbf{A}|$ stand for the rank and the determinant of the matrix $\mathbf{A}$, respectively.  By $\mathbf{A}\succeq\mathbf{0}$
or $\mathbf{A}\succ\mathbf{0}$ , it is meant that the matrix $\mathbf{A}$
is positive semidefinite or positive definite, respectively. $\mathbf{I}_{M}$ is  the identity matrix of size $M\times M$. $\mathbb{C^{N\times M}}$
denotes an $N\times M$ complex matrix. $\mathrm{diag}\left(\mathbf{A}\right)$  is a diagonal matrix with the
entries of matrix $\mathbf{A}$ as its diagonal entries.    $\left[\cdot\right]^{+}\triangleq\mathrm{max}\left\{ 0,\,\cdot\right\} $.
$|\cdot|$ and $\mathbb{E\left\{\cdot\right\}}$ denote the absolute
value and the statistical expectation, respectively. $\succeq$ represents the property of semidefinite.
$\mathcal{CN}\left(\mu,\sigma^{2}\right)$ denotes the circularly
symmetric complex Gaussian distribution with mean $\mu$ and variance
$\sigma^{2}$. $I\left(.;.\right)$ is the mutual information.\\
Additionally, a list of the most important system parameters and their meaning can be found in Table I.

\begin{table}
\caption{PARAMETERS AND THEIR MEANINGS}

$ $

\centering{}

\begin{tabular}{ll}
\hline
$\quad$$N_{s}$ & Number of Antennas of the Controller\tabularnewline
$\quad N_{i}$ & Number of Antennas of the $D_{i}$\tabularnewline
$\quad N_{e}$ & Number of Eavesdroppers\tabularnewline
$\quad\mathbf{v}_{i}$ & Transmit Beamforming Vector at Controller $S$\tabularnewline
$\quad$$s_{i}$ & Transmitted Symbol for $D_{i}$\tabularnewline
$\quad\beta$ & Power Splitting Ratio\tabularnewline
$\quad\tau$ & Transmission Time Fraction\tabularnewline
$\quad\mathbf{w}$ & Transmit Beamforming Vector at $D_{1}$\tabularnewline
$\quad\mathbf{z}$ & Artificial Noise Vector \tabularnewline
$\quad P_{s}$ & Total Transmit Power at Controller\tabularnewline
$\quad P_{t}$ & Maximum Power Transmitted at $D_{1}$ \tabularnewline
$\quad E$ & Energy Harvested by $D_{1}$\tabularnewline
$\quad R_{i}$ & Achievable Rate for Devices $D_{i}$, $i\in\left\{ 1,2\right\} $\tabularnewline
$\quad R_{sum}$ & Secrecy Sum Rate for the IoT System\tabularnewline
$\quad\gamma_{i}$ & Received SINR at $D_{i}$\tabularnewline
$\quad\mu$ & Mean of the Gaussian Noise Distribution\tabularnewline
$\quad\sigma^{2}$ & Noise Power\tabularnewline
\hline
\end{tabular}
\end{table}

\section{System model }

\begin{figure}[t]
\centering{}
\includegraphics[scale=0.7]{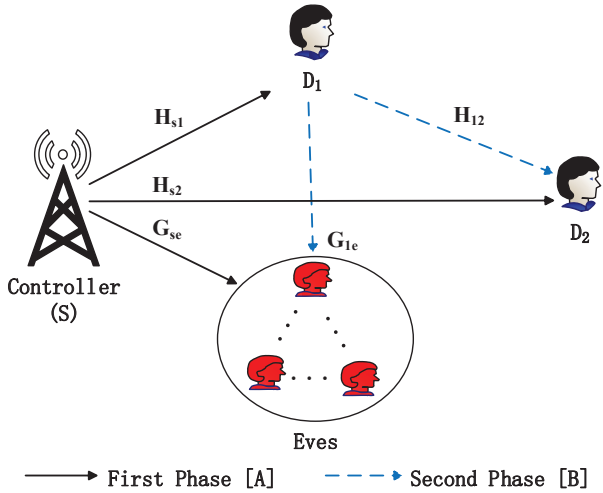}
\caption{Model for the secure cooperative IoT based communication system under consideration }
\end{figure}

As illustrated in Fig. 1,  an IoT downlink secure communication system consisting of a controller, $S$, equipped with $N_{s}$-antenna and two user devices, $D_{i}(i\in\{1,2\})$\footnote{From now on, and unless otherwise specified, the index i will always takes values from the alphabet \{1,2\}.} each equipped with $N_{i}$ antennas is considered. $S$ transmits confidential
information to the $D_{i}$  in the presence of $N_{e}$ colluding  eavesdroppers, Eves\footnote{It is noted that, although the proposed system model considers only two devices as a pair of users \cite{user_pairing}, this pairing can be generalized to multiple users following the methodology we have proposed in \cite{mayuyu}. However, this extension is beyond the scope of our current research work and as such will not be considered here.}.

It is also assumed that $D_{1}$, which operates under better channel conditions, is used for receiving rather short control signal data. $D_{2}$, which operated under weak channel conditions, which is a device which deals with some background tasks, such as downloading multimedia files. As illustrated in Fig. 1, the Eves $G_{1e}$ are located close to each other and all of them are close to S. Clearly, such configuration will lead to higher probabilities of interception for the transmitted information signal.

To ensure the high targeted secrecy rate of $D_{2}$,
$D_{1}$ acts as an EH relay to forward the information-bearing signal to  $D_{2}$. More specifically, the received signal at $D_{1}$ is split
into two parts: one for energy harvesting and the other for information decoding so that two stages will be involved in the cooperative NOMA secure transmission. At $D_{2}$ the two signals received are combined by a maximal ratio combining (MRC) diversity scheme.
It is assumed that all the fading channels shown in Fig. 1 are independent
quasi-static fading channels, which remain constant in one time slot
but vary independently from one time slot to another. Next the detailed operation of the proposed communication system will be presented.

\subsubsection*{A. Direct Transmission Phase A}

As illustrated in Fig. 1, during the $1^{st}$ stage of communication, S transmits over the channels $H_{s1}$ and $H_{s2}$ the information signals $s_1$ and $s_2$ which, after beamforming, are received by $D_1$ and $D_2$ as signals $y_1^{[A]}$ and $y_2^{[A]}$, respectively. These beamformed signals are also intercepted through the $G_{se}$ channel by the Eves, as $y_e^{[A]}$.

$D_1$ performs EH by employing a SWIPT receiver in order to perform the necessary PS. Noting that the received signals are also corrupted by independent additive white Gaussian noise (AWGN), they can be mathematically expressed as:
\begin{subequations}
\noindent
\begin{align}
 & \mathbf{\mathbf{y}}_{1}^{[A]}=\sqrt{\left(1-\beta\right)}\mathbf{H}_{s1}\left(\mathbf{v}_{1}s_{1}+\mathbf{v}_{2}s_{2}\right)+\mathbf{n}_{1}^{(1)},\label{eq:1}\\
 & \mathbf{y}_{2}^{[A]}=\mathbf{H}_{s2}\left(\mathbf{v}_{1}s_{1}+\mathbf{v}_{2}s_{2}\right)+\mathbf{n}_{2}^{(1)},\label{eq:2}\\
 & \mathbf{y}_{e}^{[A]}=\mathbf{G}_{se}\left(\mathbf{v}_{1}s_{1}+\mathbf{v}_{2}s_{2}\right)+\mathbf{n}_{e}^{(1)}.\label{eq:3}
\end{align}
\end{subequations}
In (1), $s_i$ is the transmitted information symbol for $D_i$, which has unity energy, i.e. $\mathbb{E}\left\{ |s_{i}|^{2}\right\} =1;$
$\mathbf{v}_{i}\in\mathbb{C}^{N_{s}\times1}$
is the transmit beamforming vector at $S$; $\mathbf{H}_{si}\in\mathbb{C}^{N_{i}\times N_{s}}$
and $\mathbf{G}_{se}\in\mathbb{C}^{N_{e}\times N_{s}}$ are the channel
responses from $S$ to $D_{i}$ and Eves, respectively; $\beta\in\left[0,1\right]$
is the PS ratio; $\mathbf{n}_{i}^{(1)}\sim\mathcal{CN}\left(0,\sigma_{i}^{2}\mathbf{I}_{N_{i}}\right)$
and $\mathbf{n}_{e}^{(1)}\sim\mathcal{CN}\left(0,\sigma_{e}^{2}\mathbf{I}_{N_{e}}\right)$
are the additive white complex Gaussian noise vectors at $D_{i}$ and Eves, respectively. For simplicity and without any loss of generality, it is assumed  that $\sigma_{1}^{2}=\sigma_{2}^{2}=\sigma_{e}^{2}=\sigma^{2}$.
Since the channel conditions for both $D_{1}$ and Eves are better than that of $D_{2}$, clearly $\left\Vert \mathbf{H}_{s2}\right\Vert ^{2}\leq\left\Vert \mathbf{H}_{s1}\right\Vert ^{2},$ and $\left\Vert \mathbf{H}_{s2}\right\Vert ^{2}\leq\left\Vert \mathbf{G}_{se}\right\Vert ^{2}.$
The received signals $y_1^{[A]}$ and $y_2^{[B]}$ are decoded by employing the following SIC principles. For $D_1$, as it assumed that it operates under good channel conditions, it will first detect the message $s_2$, which is intended for $D_2$, and then eliminate it from the combined signal. On the contrary, $D_2$, as it operates under bad channel conditions, it will not try to eliminate $s_1$, which is intended for $D_1$, from the combined signal.
Thus, from (\ref{eq:1}), the received signal-to-interference-plus-noise-ratio
(SINR) at $D_{1}$ for detecting $s_{2}$ can be mathematically expressed as
\noindent
\begin{align}
 & \mathrm{SINR}{}_{1,s_{2}}^{[A]}=\mathrm{\frac{\left(1-\beta\right)\left\Vert \mathbf{H}_{s1}\mathbf{v}_{2}\right\Vert ^{2}}{\left(1-\beta\right)\left\Vert \mathbf{H}_{s1}\mathbf{v}_{1}\right\Vert ^{2}+\sigma^{2}}},\label{eq:16}
\end{align}
\noindent which should be larger than or equal to a predefined threshold
so that the QoS constraint at $D_{1}$ can be satisfied. Then,for the decoding of $s_1$ at $D$, the
corresponding received signal-to-noise-ratio (SNR) is given by
\noindent
\begin{align}
 & \mathrm{SNR}{}_{1,s_{1}}^{[A]}=\frac{\left(1-\beta\right)\left\Vert \mathbf{H}_{s1}\mathbf{v}_{1}\right\Vert ^{2}}{\sigma^{2}}.\label{eq:16-1}
\end{align}
\noindent From (\ref{eq:2}), the SINR at $D_{2}$ to detect $s_{2}$ becomes
\noindent
\begin{align}
 & \mathrm{SINR}_{2,s_{2}}^{[A]}=\frac{\left\Vert \mathbf{H}_{s2}\mathbf{v}_{2}\right\Vert ^{2}}{\left\Vert \mathbf{H}_{s2}\mathbf{v}_{1}\right\Vert ^{2}+\sigma^{2}}.\label{eq:9}
\end{align}
Furthermore, the energy harvested by $D_{1}$ can be formulated as \cite{EH_PS}
\noindent
\begin{align}
 & E=\beta\left(\left\Vert \mathbf{H}_{s1}\mathbf{v}_{1}\right\Vert ^{2}+\left\Vert \mathbf{H}_{s1}\mathbf{v}_{2}\right\Vert ^{2}\right)\tau,\label{eq:9-1}
\end{align}
\noindent where $\tau=\frac{1}{2}$ is the transmission time fraction for the first phase, and the two phases have the same transmission duration.
It is assumed that the energy harvested by $D_{1}$
is mainly used for transmitting information and AN. Therefore, the
maximum power transmitted at $D_{1}$ is given by \cite{EH_rate}
\noindent
\begin{align}
 & P_{t}=\frac{E}{1-\tau}=\beta\left(\left\Vert \mathbf{H}_{s1}\mathbf{v}_{1}\right\Vert ^{2}+\left\Vert \mathbf{H}_{s1}\mathbf{v}_{2}\right\Vert ^{2}\right).\label{eq:5}
\end{align}

\subsubsection*{B. Cooperative Transmission Phase B}
During the 2nd phase of transmission,
$s_{2}$ is beamformed by $\mathbf{w}\in\mathbb{C}^{N_{1}\times1}$
and  broadcast together with AN to $D_{2}$ with the harvested energy.
Thus, the signal transmitted by $D_{1}$ can be expressed as
\noindent
\begin{align}
 & \mathbf{x}=\mathbf{w}s_{2}+\mathbf{z},\label{eq:6-2}
\end{align}
\noindent where $\mathbf{z}$ is the  AN beamforming vector generated
by $D_{1}$, whose distribution follows $\mathcal{CN}(0,\mathbf{\boldsymbol{\Sigma}})$
with $\boldsymbol{\Sigma}\succeq\mathbf{0}.$ Then, the observations
at $D_{2}$ and at the Eves can be expressed as
\begin{subequations}
\noindent
\begin{align}
 & \mathbf{y}_{2}^{[B]}=\mathbf{H}_{12}\mathbf{x}+\mathbf{n}_{2}^{(2)},\label{eq:6}\\
 & \mathbf{y}_{e}^{[B]}=\mathbf{G}_{1e}\mathbf{x}+\mathbf{n}_{e}^{(2)},\label{eq:6-1}
\end{align}
\end{subequations}
\noindent respectively, where $\mathbf{H}_{12}\in\mathbb{C}^{N_{2}\times N_{1}}$
and $\mathbf{G}_{1e}\in\mathbb{C}^{N_{e}\times N_{1}}$ are the channel
responses from $D_{1}$ to $D_{2}$ and to the Eves, respectively. $\mathbf{n}_{2}^{(2)}\sim\mathcal{CN}\left(0,\sigma^{2}\mathbf{I}_{N_{2}}\right)$
and $\mathbf{n}_{e}^{(2)}\sim\mathcal{CN}\left(0,\sigma^{2}\mathbf{I}_{N_{e}}\right)$
are the additive white complex Gaussian noise vectors at $D_{2}$
and Eves, respectively. Consequently, the SINR at $D_{2}$ to detect
$s_{2}$ can be written as
\noindent
\begin{align}
 & \mathrm{SINR}_{2,s_{2}}^{[B]}=\frac{\left\Vert \mathbf{H}_{12}\mathbf{w}\right\Vert ^{2}}{\left\Vert \mathbf{H}_{12}\mathbf{z}\right\Vert ^{2}+\sigma^{2}}.\label{eq:10}
\end{align}
Thus, the combined  SINR at $D_{2}$ becomes
\noindent
\begin{align}
 & \mathrm{SINR}_{2,s_{2}}=\mathrm{SINR}_{2,s_{2}}^{[A]}+\mathrm{SINR}_{2,s_{2}}^{[B]}.\label{eq:11}
\end{align}
\noindent As previously noted, $s_2$ is jointly detected at $D_2$ from the signals received from $S$ and $D_1$ by employing MRC diversity reception.

From (\ref{eq:3}) and (\ref{eq:6-1}), the received signal of Eves
for the two phases can be given by
\begin{align}
 & \mathbf{y}_{e}=\mathbf{H}_{e}\mathbf{s}+\mathbf{n}_{e},\label{eq:16-2-2}
\end{align}
\noindent where
\begin{subequations}
\noindent
\begin{align}
 & \mathbf{y}_{e}=\left[\begin{array}{c}
\mathbf{y}_{e}^{[A]}\\
\mathbf{y}_{e}^{[B]}
\end{array}\right],\mathbf{s}=\left[\begin{array}{c}
s_{1}\\
s_{2}
\end{array}\right],\label{eq:6-4}\\
 & \mathbf{H}_{e}=\left[\begin{array}{cc}
\mathbf{G}_{se}\mathbf{v}_{1}\, & \mathbf{G}_{se}\mathbf{v}_{2}\\
\mathbf{0}_{N_{e}\times1} & \mathbf{G}_{1e}\mathbf{w}
\end{array}\right],\label{eq:6-1-2}\\
 & \mathbf{n}_{e}=\left[\begin{array}{c}
\mathbf{n}_{e}^{(1)}\\
\mathbf{G}_{1e}\mathbf{z}+\mathbf{n}_{e}^{(2)}
\end{array}\right].\label{eq:6-1-1}
\end{align}
\end{subequations}

\noindent From (\ref{eq:16-2-2}), the information sum rate leaked
to the Eves can be written as \cite{Eve_secrecyrate}
\begin{subequations}
\begin{align}
 & R_{e}=I\left(s_{i};\mathbf{y}_{e}\right)=\frac{1}{2}\mathrm{log}\left|\mathbf{I}_{N_{e}}+\mathbf{H}_{e}\mathbf{H}_{e}^{H}\mathbf{Q}_{e}^{-1}\right|=\mathrm{\frac{1}{2}log}\nonumber \\
 & \left(1+\frac{\mathrm{Tr\left(\mathbf{G}_{se}\left(\stackrel[i=1]{2}{\sum}\mathbf{v}_{\mathit{i}}\mathbf{v}_{\mathit{i}}^{H}\right)\mathbf{G}_{se}^{H}+\mathbf{G}_{1e}\mathbf{w}\mathbf{w}^{H}\mathbf{G}_{1e}^{H}\right)}}{\mathrm{Tr}\left(\mathbf{Q}_{e}\right)}\right),
\end{align}
\end{subequations}
\noindent where $\mathbf{Q}_{e}=\mathbf{n}_{e}\mathbf{n}_{e}^{H}=\mathrm{diag}\left(\sigma^{2}\mathbf{I}_{N_{e}},\:\sigma^{2}\mathbf{I}_{N_{e}}+\mathbf{G}_{1e}\boldsymbol{\Sigma}\mathbf{G}_{1e}^{H}\right)$,
$i\in\left\{ 1,2\right\} ,$  and the factor $\frac{1}{2}$
is introduced since the messages are transmitted in two consecutive phases \cite{Eves_receive}. Note that the minimum-mean-square-error criterion and SIC have been employed by Eves. The achievable rates of $D_{1}$
and $D_{2}$ are given by
\begin{subequations}
\begin{align}
 & R_{1}=I\left(s_{1};\mathbf{y}_{1}\right)=\frac{1}{2}\mathrm{log}\left(1+\mathrm{SNR_{1,s_{1}}}\right),\label{eq:16-2-4}\\
 & R_{2}=I\left(s_{2};\mathbf{y}_{2}\right)=\frac{1}{2}\mathrm{log}\left(1+\mathrm{SINR_{2,s_{2}}}\right),\label{eq:16-2}
\end{align}
\end{subequations}
\noindent respectively. As pointed out it \cite{Eves_receive}, it is very difficult to obtain the secrecy capacity region of the downlink
wireless systems. Alternative, the rate difference between the legitimate sum rate and
the information sum rate leaked to the Eves (i.e., $R_{e}$) can be computed as the secure
performance measure. In this case, the SSR can be expressed as \cite{Secrecy_sum_rate}
\begin{align}
 & R_{sum}=\left[\stackrel[i=1]{2}{\sum}R_{i}-R_{e}\right]^{+}.\label{eq:16-1-2}
\end{align}

To improve the security of the downlink for the cooperative communication system under consideration, an SSR maximization
problem is formulated.
The transmit power constraint at $S$,
the SIC constraint at $D_{1}$  and  the EH constraint should be satisfied in the SSR maximization
problem. The EH constraint is that the  achievable
power for transmitting information bearing signal and AN is smaller
than or equal to the energy harvested by $D_{1}$.  According to the
result given in \cite{fairness}, a higher sum rate can be obtained
if $S$ allocates more power for the information-bearing signal $s_{1}$.
However, such an approach is not applicable in the optimization problem consider here for which the user fairness and high targeted secrecy rate at $D_{2}$  should be ensured \cite{fairness}. With the above in mind, the optimization problem considered in this paper can be formulated as

\begin{subequations}\label{max_1}
\begin{align}
\underset{\mathbf{v}_{1},\mathbf{v}_{2}}{\mathrm{max}}\,\: & R_{sum}\label{eq:17a}\\
\mathrm{s.t.\,\;\:} & \mathrm{SINR}{}_{1,s_{2}}\geq\gamma,\,\mathrm{SINR}_{2,s_{2}}\geq\gamma,\label{eq:17e}\\
 & \left\Vert \mathbf{v}_{1}\right\Vert ^{2}+\left\Vert \mathbf{v}_{2}\right\Vert ^{2}\leq P_{s},\left\Vert \mathbf{w}\right\Vert ^{2}+\left\Vert \mathbf{z}\right\Vert ^{2}\leq P_{t},\label{eq:17f}
\end{align}
\end{subequations}
where $P_{s}$ is the transmit power constraint at $S$,
$\left\Vert \mathbf{w}\right\Vert ^{2}$ and $\left\Vert \mathbf{z}\right\Vert ^{2}$
are the allocated power for information transmission and AN, respectively.
Note that (\ref{eq:17e}) indicates that the received SINR to decode $s_{2}$ should
be no less than the SINR threshold $\gamma$, as (\ref{eq:17e}) relates to the SIC constraint at $D_1$ to ensure the necessary QoS decoding performance of $D_2$.

Next the NP-hard problem of (\ref{max_1}) will be solved by optimizing $\mathbf{v}_{1},\mathbf{v}_{2}$, $\mathbf{w}$ as well as $\beta$ for the cases of passive and active Eves (see Sections III and IV, respectively).
\section{AN-aided secure beamforming design (ASBD) for the passive eavesdroppers case }
In this section, it is assumed that the passive Eves' CSI is not available at the
transmitters $D_{1}$ and $S$. For example this occurs when the eavesdropping nodes in IoT systems are passive
or malicious. Since in this case even the location of the Eves is hard to obtain \cite{ECSI}, and thus obtaining CSI will be even more difficult, a AN-aided beamforming design scheme is proposed.

\subsubsection*{Zero-Forcing (ZF) Constraint on AN}

Since $D_{1}$ has no CSI for the   Eves, i.e., $D_{1}$ doesn't know $\mathbf{G}_{1e}$,
in order to eliminate the interference at $D_{2}$, a ZF condition on the
AN beamforming is applied \cite{my_paper,AN_number}. In other words, the AN beamforming is designed so that it steers the signal into the null space of $\mathbf{H}_{12}$, so that $\mathbf{H}_{12}\mathbf{z}=\mathbf{0}$.
In this way, $D_{1}$ cannot transmit AN to interfere
the Eves selectively under the Zero Forcing constraint \cite{non_colluding}. Hence, the AN $\mathbf{z}$ is in the form of $\mathbf{z}=\mathbf{H}_{\perp}\mathbf{n},$ where $\mathbf{H}_{\perp}=\mathbf{I}_{N_{2}}-\mathbf{H}_{12}\left(\mathbf{H}_{12}^{H}\mathbf{H}_{12}\right)^{-1}\mathbf{H}_{12}^{H}$
is the projection matrix onto the null space of $\mathbf{H}_{12}$.
Note that  the components of $\mathbf{n}$ are independent and identically distributed (iid) Gaussian variables, which follow $\mathcal{CN}\left(0,\sigma_{z}^{2}\right)$ so that the spatial covariance of AN can be expressed
as $\boldsymbol{\Sigma}=\sigma_{z}^{2}\mathbf{H}_{\perp}\mathbf{H}_{\perp}^{H}\succeq\mathbf{0}$.

The SSR $R_{sum}$ is maximized only when $R_{e}$ is minimized.
Thus, in order to minimize $R_{e}$, the power allocated for AN should be made  as much as possible,  under the EH and SIC constraints at $D_{1}$,
the transmit power constraint at $S$ and QoS constraint at $D_{i}$. According to the semidefinite
relaxation (SDR) technique, let $\mathbf{V}_{i}=\mathbf{v}_{i}\mathbf{v}_{i}^{H},\,i\in\left\{ 1,2\right\} ,\mathbf{W}=\mathbf{w}\mathbf{w}^{H}$
and drop rank-one constraints $\mathrm{rank}\left(\mathbf{V}_{i}\right)=1,\mathrm{\,rank}\left(\mathbf{W}\right)=1.$
Thus, the secure beamforming problem is reformulated as
\begin{subequations}\label{max_2}
\noindent
\begin{align}
\underset{\mathbf{V}_{i}\succeq\mathbf{0},\mathbf{W}\succeq\mathbf{0},\,\beta}{\mathrm{max}}\, & \mathrm{Tr}\left(\boldsymbol{\Sigma}\right)\label{eq:17a-1-2}\\
\mathrm{s.t.\quad\;\:\;\,} & \mathrm{\left(1+\frac{\left(1-\beta\right)\mathrm{Tr}\left(\mathbf{W}_{1}^{1}\right)}{\sigma^{2}}\right)\geq\mathit{r}_{1},}\label{eq:17b-1-2}\\
 & \mathrm{\left(1+\frac{\mathrm{Tr}\left(\mathbf{W}_{2}^{2}\right)}{\mathrm{Tr}\left(\mathbf{W}_{2}^{1}\right)+\sigma^{2}}+\frac{\mathrm{Tr}\left(\mathbf{W}_{12}\right)}{\sigma^{2}}\right)\geq\mathit{r}_{2},}\label{eq:17b-1-3}\\
 & \frac{\left(1-\beta\right)\mathrm{\mathrm{T}r}\left(\mathbf{W}_{1}^{2}\right)}{\sigma^{2}+\left(1-\beta\right)\mathrm{Tr}\left(\mathbf{W}_{1}^{1}\right)}\geq\gamma,\label{eq:17d-1-2}\\
 & \frac{\mathrm{Tr}\left(\mathbf{W}_{2}^{2}\right)}{\mathrm{Tr}\left(\mathbf{W}_{2}^{1}\right)+\sigma^{2}}+\frac{\mathrm{Tr}\left(\mathbf{W}_{12}\right)}{\sigma^{2}}\geq\gamma,\label{eq:17d-1-3}\\
 & \mathrm{Tr}\left(\mathbf{W}\right)+\mathrm{Tr}\left(\boldsymbol{\Sigma}\right)\leq\beta\mathrm{Tr}\left(\stackrel[i=1]{2}{\sum}\mathbf{W}_{1}^{i}\right),\label{eq:17g-1-1}\\
 & \mathrm{Tr}\left(\mathbf{V}_{1}\right)+\mathrm{Tr}\left(\mathbf{V}_{2}\right)\leq P_{s}.\label{eq:13}
\end{align}
\end{subequations}
\noindent where $\mathbf{W}_{1}^{i}=\mathbf{H}_{s1}\mathbf{V}_{i}\mathbf{H}_{s1}^{H},$
$\mathbf{W}_{2}^{i}=\mathbf{H}_{s2}\mathbf{V}_{i}\mathbf{H}_{s2}^{H},\,i\in\left\{ 1,2\right\} $,
$\mathbf{W}_{12}=\mathbf{H}_{12}\mathbf{W}\mathbf{H}_{12}^{H}$, $r_{i}$
is the predefined threshold of achievable ergodic rate of $D_{i}$,
and $\frac{1}{2}\mathrm{log}(\cdot)$ is not included in (\ref{eq:17b-1-2})
and (\ref{eq:17b-1-3}) since the logarithmic function is a monotonically
increasing function. It is noted that the optimization problem (\ref{max_2})
is non-convex due to the non-convex constraints (\ref{eq:17b-1-2})-(\ref{eq:17g-1-1}).
Consequently, these non-convex constraints will be transformed into equivalent convex forms based on the idea of SCA, which can be used to iteratively approximate a non-convex optimization problems by an equivalent convex problem \cite{SCA_my}.

\subsubsection*{Transformation of (\ref{eq:17b-1-2})}

Based on the SCA method and the ER,
(\ref{eq:17b-1-2}) can be replaced by the following
two constraints
\begin{subequations}
\noindent
\begin{align}
 & \theta_{1}^{2}\geq\left(r_{1}-1\right)\sigma^{2},\label{eq:16a-1-1-1}\\
 & \left[\begin{array}{cc}
(1-\beta) & \theta_{1}\\
\theta_{1} & \mathrm{Tr}(\mathbf{W}_{1}^{1})
\end{array}\right]\succeq\mathbf{0}.\label{eq:16b-1-1-1}
\end{align}
\end{subequations}

\begin{lemma}
For a standard convex function $g(x)$ and a concave function $f(x)$,
if the constraint   $g(x)\leq f(x)$  exists, then the constraint is convex \cite{djk}.
\end{lemma}

Based upon the following Lemma, it is clear that (\ref{eq:16a-1-1-1}) is non-convex. Therefore, the first order Taylor expansion (FOTE) can be utilized to approximate (\ref{eq:16a-1-1-1}) with the following convex function
\noindent
\begin{align}
 & 2\theta_{1}^{(n)}\theta_{1}-(\theta_{1}^{(n)})^{2}\geq\left(r_{1}-1\right)\sigma^{2},\label{eq:16-1-2-1-1-1}
\end{align}
\noindent where $\theta_{1}^{(n)}$ is the value of variable $\theta_{1}$
at the $n$-th iteration, i.e. the superscript $(n)$ denotes the point obtained during the $n$-th iteration. Note that the  constraint (\ref{eq:17b-1-2})
is now replaced by two convex constraints, i.e., the linear matrix inequality
constraint (\ref{eq:16b-1-1-1}) and the convex constraint (\ref{eq:16-1-2-1-1-1}).

\subsubsection*{Transformation of (\ref{eq:17b-1-3})}

Similarly, based on  ER and FOTE, the non-convex constraint
(\ref{eq:17b-1-3}) can be approximated by the following convex functions
\begin{subequations}\label{SCA1}
\begin{align}
 & 2\psi_{1}^{(n)}\psi_{1}-\left(\psi_{1}^{(n)}\right)^{2}\geq D+\sigma^{2}\left(r_{2}-1\right),\\
 & \left[\begin{array}{cc}
\frac{\mathrm{Tr}\left(\mathbf{W}_{12}\right)}{\sigma^{2}} & \psi_{1}\\
\psi_{1} & \mathrm{Tr}\left(\mathbf{W}_{2}^{1}\right)
\end{array}\right]\succeq\mathbf{0},
\end{align}

\end{subequations}

\noindent where $D=\mathrm{Tr}\left(\mathit{r}_{2}\mathbf{W}_{2}^{1}-\mathbf{W}_{2}^{1}-\mathbf{W}_{2}^{2}-\mathbf{W}_{12}\right)$.

\subsubsection*{Transformation of (\ref{eq:17d-1-2})}

The non-convex constraint (\ref{eq:17d-1-2}) is equivalently formulated
as
\noindent
\begin{align}
 & \left(1-\beta\right)\mathrm{\mathrm{T}r}\left(\mathbf{W}_{1}^{2}-\gamma\mathbf{W}_{1}^{1}\right)\geq\gamma\sigma^{2}.\label{eq:6-3}
\end{align}

\noindent  As (\ref{eq:6-3}) is still a non-convex constraint, ER and FOTE are employed to yield the following  approximate convex constraint
\begin{subequations}\label{eq:15-1}

\noindent
\begin{align}
 & 2\phi_{1}^{(n)}\phi_{1}-(\phi_{1}^{(n)})^{2}\geq\gamma\sigma^{2},\label{eq:9-2}\\
 & \left[\begin{array}{cc}
\left(1-\beta\right) & \phi_{1}\\
\phi_{1} & \mathrm{\mathrm{T}r}\left(\mathbf{W}_{1}^{2}-\gamma\mathbf{W}_{1}^{1}\right)
\end{array}\right]\succeq\mathbf{0}.\label{eq:10-1}
\end{align}

\end{subequations}

\subsubsection*{Transformation of (\ref{eq:17d-1-3})}

Similarly, the non-convex constraint (\ref{eq:17d-1-3})
is converted to the following convex form
\begin{subequations}\label{eq:15-1-1}

\noindent
\begin{align}
 & 2\omega_{1}^{(n)}\omega_{1}-(\omega_{1}^{(n)})^{2}\geq\mathrm{\mathrm{T}r}\left(\gamma\mathbf{W}_{2}^{1}-\mathbf{W}_{12}-\mathbf{W}_{2}^{2}\right)+\sigma^{2}\gamma,\label{eq:20}\\
 & \left[\begin{array}{cc}
\frac{\mathrm{Tr}\left(\mathbf{W}_{12}\right)}{\sigma^{2}} & \omega_{1}\\
\omega_{1} & \mathrm{\mathrm{T}r}\left(\mathbf{W}_{2}^{1}\right)
\end{array}\right]\succeq\mathbf{0}.\label{eq:7}
\end{align}

\end{subequations}

\subsubsection*{Transformation of (\ref{eq:17g-1-1})}

In the same manner,  the non-convex constraint (\ref{eq:17g-1-1}) can be approximated by
\begin{subequations}\label{eq:15}

\noindent
\begin{align}
 & 2\varphi_{1}^{(n)}\varphi_{1}-(\varphi_{1}^{(n)})^{2}\geq\mathrm{Tr}\left(\mathbf{W}\right)+\mathrm{Tr}\left(\boldsymbol{\Sigma}\right),\label{eq:9-2-1}\\
 & \left[\begin{array}{cc}
\beta & \varphi_{1}\\
\varphi_{1} & \mathrm{Tr}\left(\stackrel[i=1]{2}{\sum}\mathbf{W}_{1}^{i}\right)
\end{array}\right]\succeq\mathbf{0}.\label{eq:10-1-1}
\end{align}

\end{subequations}

Thus, the optimization problem (\ref{max_2}) has been  recast as
\begin{subequations}\label{max_3-1}

\noindent
\begin{align}
\underset{\begin{array}{c}
\begin{array}{c}
\mathbf{V}_{i},\mathbf{W},\,\beta,\theta_{1},\\
\psi_{1},\phi_{1},\omega_{1},\varphi_{1}
\end{array}\end{array}}{\mathrm{max}} & \mathrm{Tr}\left(\boldsymbol{\Sigma}\right)\label{eq:25a-1}\\
\mathrm{s.t.\quad\quad\quad\:} & (\ref{eq:13}),(\ref{eq:16b-1-1-1}),(\ref{eq:16-1-2-1-1-1}),\label{eq:25b-1}\\
 & (\ref{SCA1}),(\ref{eq:15-1})-(\ref{eq:15}),\label{eq:25c-1}\\
 & \mathbf{V}_{i}\succeq\mathbf{0},\mathbf{W}\succeq\mathbf{0},\,i\in\left\{ 1,2\right\} ,\label{eq:30}
\end{align}

\end{subequations}

\noindent which is a semidefinite programming and its solution
$(\mathbf{V}_{1}^{*},\mathbf{V}_{2}^{*},\mathbf{W}^{*},\boldsymbol{\Sigma}^{*})$
can be obtained by off-the-shelf optimization solvers, e.g., SeDuMi
or Yalmip\cite{Yalmip}.

The convex problem (\ref{max_3-1}) is solved in an iterative manner starting with the initial values of
$\left(\theta_{1}^{\left(0\right)},\psi_{1}^{\left(0\right)},\phi_{1}^{\left(0\right)},\omega_{1}^{\left(0\right)},\varphi_{1}^{(0)}\right),$
as presented  in Algorithm 1. During each iteration, $\left(\theta_{1}^{\left(n\right)},\psi_{1}^{\left(n\right)},\phi_{1}^{\left(n\right)},\omega_{1}^{\left(n\right)},\varphi_{1}^{(n)}\right)$
is updated based on the previously obtained solution until the SSR
gap between two successive iterations is less than a predefined accuracy
$\epsilon_{1}$. Moreover,  based on Lemma 3.1 of \cite{rank_one},  the solution $(\mathbf{V}_{1}^{*},\mathbf{V}_{2}^{*},\mathbf{W}^{*})$
yielded by SDR is rank-one.
Then the beamforming vectors $\mathbf{v}_{1},\mathbf{v}_{2}$ and $\mathbf{w}$
are obtained by eigenvalue decomposition of $\mathbf{V}_{1}, \mathbf{V}_{2}$ and $\mathbf{W}$, respectively.

\begin{algorithm}[t]
\caption{Proposed ASBD Scheme}

1: \textbf{Input:} Set $n=0$, $\theta_{1}^{(0)}=1$, $\psi_{1}^{(0)}=1,\,\phi_{1}^{(0)}=1,$

$\quad\,\omega_{1}^{(0)}=1,$$\varphi_{1}^{(0)}=1,$ $\delta=1$,
and $\epsilon_{1}=10^{-4}$.

2: \textbf{while $\delta\geq\epsilon_{1}$ do}

a) Solve (\ref{max_3-1}) and update $\left(\theta_{1}^{\left(n\right)},\psi_{1}^{\left(n\right)},\phi_{1}^{\left(n\right)},\omega_{1}^{\left(n\right)},\varphi_{1}^{(n)}\right)$$\leftarrow$

$\quad\,$$(\theta_{1}^{*},\psi_{1}^{*},\phi_{1}^{*},\omega_{1}^{*},\varphi_{1}^{*})$,

b) Update $\beta^{(n)},\mathbf{V}_{i}^{(n)},\mathbf{W}^{(n)}$ and
$\boldsymbol{\Sigma}^{(n)}$.

c) Update $\delta=\left|\mathrm{Tr}\left(\boldsymbol{\Sigma}^{(n)}\right)-\mathrm{Tr}\left(\boldsymbol{\Sigma}^{(n-1)}\right)\right|.$

d) Set $n\leftarrow n+1$.

3: \textbf{end while}

4: \textbf{Output: $\beta^{*},\mathbf{V}_{i}^{*},$ $\mathbf{W}^{*}$
}and\textbf{ $\boldsymbol{\Sigma}^{*}.$}
\end{algorithm}

Finally,   the information rate leaked to Eves can be shown in (\ref{eq:18}) (in the top of the next page). Substituting (\ref{eq:18}) into (\ref{eq:16-1-2})  and replacing $(\mathbf{v}_{1},\mathbf{v}_{2},\mathbf{w})$ with $(\mathbf{v}_{1}^{*},\mathbf{v}_{2}^{*},\mathbf{w}^{*})$,
the SSR of the IoT system under consideration can be computed.

\begin{algorithm*}[t]

\begin{align}\label{eq:18}
  R_{e} & =\mathrm{\frac{1}{2}log}\left(1+\frac{\mathrm{Tr\left(\mathbf{G}_{se}\left(\stackrel[i=1]{2}{\sum}\mathbf{v}_{\mathit{i}}^{*}\mathbf{v}_{\mathit{i}}^{*H}\right)\mathbf{G}_{se}^{H}+\mathbf{G}_{1e}\mathbf{w^{*}}\mathbf{w}^{*H}\mathbf{G}_{1e}^{H}\right)}}{\mathrm{Tr}\left(\mathrm{diag}\left(\sigma^{2}\mathbf{I}_{N_{e}},\:\sigma^{2}\mathbf{I}_{N_{e}}+\mathbf{G}_{1e}\boldsymbol{\Sigma}^{*}\mathbf{G}_{1e}^{H}\right)\right)}\right).
\end{align}

\end{algorithm*}

\section{Secure beamforming design for the active eavesdroppers case }

When active Eves are present, their CSI can be estimated by $S$ and $D_1$ through leakage from the Eves' receiver radio frequency (RF) frontend \cite{ECSI}. Thus, the AN-aided transmission beamforming design proposed in Section III may not be the best solution because of the presence of active Eves. In light of this, an orthogonal-projection-based transmission beamforming design is proposed for the SSR maximization problem defined in (\ref{max_1}). For this scheme it is noted that the  number of antennas at $D_{1}$ should be more
than that of the Eves (i.e., $N_{1}>N_{e}$). For the case when $S$ and $D_i$ have single antennas, a simple power control scheme which optimizes the SSR is also proposed.

\subsection*{A. Multi-Antennas Configuration}

The main idea behind the new approach can be explained as follows. Firstly, by using the available CSI of the Eves, a new design for the transmit beamforming $\mathbf{w}$ is proposed, which will deteriorate the quality of the signals received by the Eves due to the addition of AN. This design will be termed as Orthogonal-Projection-Based Secure Beamforming Design (OSBD) for Multi-Antennas Case. On that basis, by letting AN $\mathbf{z}=\mathbf{0}$, the transmitted signal
from $D_{1}$ during the second phase becomes
\noindent
\begin{align}
 & \mathbf{x}=\mathbf{w}s_{2}.
\end{align}

\noindent The observations at $D_{2}$ and Eves can be expressed as

\begin{subequations}

\noindent
\begin{align}
 & \mathbf{y}_{2}^{[B]}=\mathbf{H}_{12}\mathbf{w}s_{2}+\mathbf{n}_{2}^{(2)},\label{eq:6-5}\\
 & \mathbf{y}_{e}^{[B]}=\mathbf{G}_{1e}\mathbf{w}s_{2}+\mathbf{n}_{e}^{(2)},\label{eq:6-1-3}
\end{align}

\end{subequations}

\noindent respectively, where  $\mathbf{w}$  lies in the null space of Eves' channel $\mathbf{G}_{1e}$,
i.e., $\mathbf{G}_{1e}\mathbf{w}=\mathbf{0}$.  Therefore, $\mathbf{w}\mathbf{w}^{H}=\alpha\left(\frac{\mathbf{I}_{N_{1}}-\mathbf{Q}_{1e}}{\left\Vert \mathbf{I}_{N_{1}}-\mathbf{Q}_{1e}\right\Vert }\right)$
and $\mathbf{Q}_{1e}=\mathbf{G}_{1e}^{H}(\mathbf{G}_{1e}^{H}\mathbf{G}_{1e})^{-1}\mathbf{G}_{1e}$, where
$\alpha\in\left[0,1\right]$ is a scale factor determining the power invested on transmit
beamforming.  The received SINRs at $D_{2}$ and Eves are given by
\begin{subequations}

\noindent
\begin{align}
 & \mathrm{SINR_{2,s_{2}}}=\frac{\left\Vert \mathbf{H}_{s2}\mathbf{v}_{2}\right\Vert ^{2}}{\left\Vert \mathbf{H}_{s2}\mathbf{v}_{1}\right\Vert ^{2}+\sigma^{2}}+\frac{\left\Vert \mathbf{H}_{12}\mathbf{w}\right\Vert ^{2}}{\sigma^{2}},\label{eq:16a-2}\\
 & \mathrm{SINR_{\mathit{e}}}=\frac{\mathrm{Tr\left(\mathbf{G}_{se}\left(\stackrel[i=1]{2}{\sum}\mathbf{v}_{\mathit{i}}\mathbf{v}_{\mathit{i}}^{H}\right)\mathbf{G}_{se}^{H}\right)}}{\mathrm{Tr}\left(\mathbf{\Theta}\right)},\label{eq:16b-1-2}
\end{align}

\noindent \end{subequations}

\noindent respectively, where $\mathbf{\Theta}=\mathrm{diag}\left(\sigma^{2}\mathbf{I}_{N_{e}},\:\sigma^{2}\mathbf{I}_{N_{e}}\right)$.

To solve the optimization problem (\ref{max_1}), we first let $\mathbf{V}_{i}=\mathbf{v}_{i}\mathbf{v}_{i}^{H},i\in\{1,2\},\,\mathbf{W}=\mathbf{w}\mathbf{w}^{H},$
and drop rank-one constraints $\mathrm{rank}\left(\mathbf{V_{\mathit{i}}}\right)=1,\,\mathrm{rank}\left(\mathbf{W}\right)=1.$
Then, (\ref{max_1}) can be reformulated as (\ref{max_1-1}) (on the top of next page), where $\mathbf{W}_{e}^{i}=\mathbf{G}_{se}\mathbf{V}_{i}\mathbf{G}_{se}^{H}$, and the non-constraints (\ref{eq:17d-1-2}) and (\ref{eq:17d-1-3})
have been converted to convex forms as previously explained in Section III. It is noted that (\ref{max_1-1}) is non-convex due to the non-convex
objective function (\ref{eq:17a-1}) and the constraint (\ref{eq:17b-1}).
However, by using FOTE and ER,
(\ref{eq:17b-1}) can be approximated by the following convex constraint

\begin{algorithm*}[tbh]

\begin{subequations}\label{max_1-1}
\noindent

\begin{align}
\underset{\mathbf{V}_{1}\succeq\mathbf{0},\mathbf{V}_{2}\succeq\mathbf{0},\mathbf{W}\succeq\mathbf{0}}{\mathrm{max}}\,\, & \frac{1}{2}\mathrm{log}\left(\left(1+\frac{\left(1-\beta\right)\mathrm{Tr}\left(\mathbf{W}_{1}^{1}\right)}{\mathrm{Tr}\left(\mathbf{\Theta}\right)}\right)\left(1+\frac{\mathrm{Tr}\left(\mathbf{W}_{2}^{2}\right)}{\sigma^{2}+\mathrm{Tr}\left(\mathbf{W}_{2}^{1}\right)}+\frac{\mathrm{Tr}\left(\mathbf{W}_{12}\right)}{\sigma^{2}}\right)\left(\frac{\mathrm{Tr}\left(\mathbf{\Theta}\right)}{\mathrm{Tr}\left(\mathbf{\Theta}+\stackrel[i=1]{2}{\sum}\mathbf{W}_{e}^{i}\right)}\right)\right)\label{eq:17a-1}\\
\mathrm{s.t.\quad\quad\:\:\,\,\;} & \mathrm{Tr}(\mathbf{W})\leq P_{t},\label{eq:17b-1}\\
 & (\ref{eq:17d-1-2}),(\ref{eq:17d-1-3}),\,\mathrm{and}\,(\ref{eq:13}),\label{eq:17c-1}
\end{align}

\end{subequations}

\end{algorithm*}

\begin{subequations}\label{FOTE}
\noindent
\begin{align}
 & 2\varphi_{2}^{(n)}\varphi_{2}-(\varphi_{2}^{(n)})^{2}\geq\mathrm{Tr}\left(\mathbf{W}\right),\label{eq:11-2-1}\\
 & \left[\begin{array}{cc}
\beta & \varphi_{2}\\
\varphi_{2} & \mathrm{Tr}\left(\stackrel[i=1]{2}{\sum}\mathbf{W}_{1}^{i}\right)
\end{array}\right]\succeq\mathbf{0}.\label{eq:12-1}
\end{align}
\end{subequations}\noindent Since problem (\ref{max_1-1}) is still non-convex its objective function will be transformed into the convex form based on the SCA method. For this, by introducing the slack
variables $\tau$, $\mu_{i}\geq0,i\in\{1,2\},$ the objective function
can be alternatively reformulated as
\begin{subequations}\label{max_3-2}

\noindent
\begin{align}
\underset{\begin{array}{c}
\tau,\mu_{i}\geq0,\beta\end{array}}{\mathrm{max}} & \stackrel[i=1]{2}{\sum}\mu_{i}-\tau\label{eq:25a-2}\\
\mathrm{s.t.\quad\quad\:} & \mathrm{log}\left(1+\frac{\left(1-\beta\right)\mathrm{Tr}\left(\mathbf{W}_{1}^{1}\right)}{\mathrm{Tr}\left(\mathbf{\Theta}\right)}\right)\geq\mu_{1},\label{eq:25b-2}\\
 & \mathrm{log}\left(1+\frac{\mathrm{Tr}\left(\mathbf{W}_{2}^{2}\right)}{\sigma^{2}+\mathrm{Tr}\left(\mathbf{W}_{2}^{1}\right)}+\frac{\mathrm{Tr}\left(\mathbf{W}_{12}\right)}{\sigma^{2}}\right)\geq\mu_{2},\label{eq:25c-2}\\
 & \mathrm{log}\left(\frac{\mathrm{Tr}\left(\mathbf{\Theta}\right)}{\mathrm{Tr}\left(\mathbf{\Theta}+\stackrel[i=1]{2}{\sum}\mathbf{W}_{e}^{i}\right)}\right)\leq\tau,\label{eq:21}
\end{align}

\end{subequations}

\noindent where factor $\frac{1}{2}$  in (\ref{max_1-1}) is omitted
since it does not affect the monotonicity of  (\ref{max_1-1}).
Then, the SCA can be implemented over (\ref{max_3-2}). It
can be seen that (\ref{eq:25b-2}) is equivalent to
\noindent
\begin{align}
 & \left(1-\beta\right)\mathrm{Tr}\left(\mathbf{W}_{1}^{1}\right)\geq\left(e^{\mu_{1}}-1\right)\mathrm{Tr}\left(\mathbf{\Theta}\right),\label{eq:16-1-2-1}
\end{align}

\noindent by the FOTE, (\ref{eq:16-1-2-1}) is transformed into

\noindent
\begin{align}
 & \left(1-\beta\right)\mathrm{Tr}\left(\mathbf{W}_{1}^{1}\right)\geq\left(x-1\right)\mathrm{Tr}\left(\mathbf{\Theta}\right),\label{eq:16a-1}
\end{align}

\noindent where $x=e^{\mu_{1}^{(n)}}(\mu_{1}-\mu_{1}^{(n)}+1)$ is
the FOTE of $e^{\mu_{1}}$ around the point $\mu_{1}^{(n)}.$ According to the ER method, (\ref{eq:16a-1}) is transformed
into the following convex form
\begin{subequations}\label{FOTE1}
\noindent
\begin{align}
 & 2\xi^{(n)}\xi-(\xi^{(n)})^{2}\geq\left(x-1\right)\mathrm{Tr}\left(\mathbf{\Theta}\right),\label{eq:22-1}\\
 & \left[\begin{array}{cc}
1-\beta & \xi\\
\xi & \mathrm{Tr}\left(\mathbf{W}_{1}^{1}\right)
\end{array}\right]\succeq\mathbf{0}.\label{eq:23-1}
\end{align}

\end{subequations}

\noindent (\ref{eq:25c-2}) is non-convex and  it  is not suitable to  handle the non-convex problem by adopting the ER method. Therefore, Proposition 1 is introduced to
convert (\ref{eq:25c-2}) into a convex form.
\begin{prop}
For any non-negative variables $x,\,y,\,z$, a non-convex expression
$xy\leq z$ can be approximated by the following convex constraint
\begin{equation}
\left(\eta x\right){}^{2}+\left(y/\eta\right){}^{2}\leq2z.\label{eq:14}
\end{equation}
\end{prop}
\begin{IEEEproof}
According to the arithmetic-geometric
mean (AGM) inequality, the approximation of the expression
$xy\leq z$ can be expressed as
\begin{equation}
2xy\leq\left(\eta x\right){}^{2}+\left(y/\eta\right){}^{2}\leq2z,\label{eq:14my}
\end{equation}

\noindent where the former inequality holds with equality, if and only
if,  $\eta=\sqrt{y/x}$. Then the  non-convex expression
$xy\leq z$ can be replaced by
$\frac{\left(\eta x\right){}^{2}+\left(y/\eta\right){}^{2}}{2}\leq z.$
It can be observed that  $\left(\eta x\right){}^{2}+\left(y/\eta\right){}^{2}\leq2z$
is a convex constraint based on Lemma 1, which completes the proof.
\end{IEEEproof}

Firstly, (\ref{eq:25c-2}) can be converted to
\begin{align}
 & \left(e^{\mu_{2}}-\frac{\mathrm{Tr}\left(\mathbf{W}_{12}\right)}{\sigma^{2}}\right)\mathrm{Tr}\left(\mathbf{W}_{2}^{1}\right)\leq\Xi,\label{eq:24-1}
\end{align}

\noindent where $\Xi=\mathrm{Tr}\left(\mathbf{W}_{2}^{2}+\mathbf{W}_{12}+\mathbf{W}_{2}^{1}\right)-e^{\mu_{2}}+\sigma^{2}$.
Using the FOTE, (\ref{eq:24-1}) is then transformed into
\begin{align}
 & \left(y-\frac{\mathrm{Tr}\left(\mathbf{W}_{12}\right)}{\sigma^{2}}\right)\mathrm{Tr}\left(\mathbf{W}_{2}^{1}\right)\leq\Xi_{1},\label{eq:25}
\end{align}

\noindent where $\Xi_{1}=\mathrm{Tr}\left(\mathbf{W}_{2}^{2}+\mathbf{W}_{12}+\mathbf{W}_{2}^{1}\right)-y+\sigma^{2}$
, $y=e^{\mu_{2}^{(n)}}(\mu_{2}-\mu_{2}^{(n)}+1)$.  Finally, according to the Proposition 1, (\ref{eq:25}) can be approximated by the following
constraint
\begin{align}
 & \left(\eta^{\left(n\right)}\left(y-\frac{\mathrm{Tr}\left(\mathbf{W}_{12}\right)}{\sigma^{2}}\right)\right)^{2}+\left(\mathrm{Tr}\left(\mathbf{W}_{2}^{1}\right)/\eta^{\left(n\right)}\right)^{2}\leq2\Xi_{1},\label{eq:26-1}
\end{align}

\noindent where $\eta^{(n)}$  can be updated by
\begin{align}
 & \eta^{\left(n\right)}=\sqrt{\left(\mathrm{Tr}\left(\mathbf{W}_{2}^{1}\right)\right)^{\left(n-1\right)}/\left(y-\frac{\mathrm{Tr}\left(\mathbf{W}_{12}\right)}{\sigma^{2}}\right)^{\left(n-1\right)}}.\label{eq:23}
\end{align}

\noindent It is clear from Lemma 1 that now (\ref{eq:26-1}) is a convex constraint.

Employing FOTE and ER, we transform constraint
(\ref{eq:21}) as follows
\begin{subequations}\label{FOTE1-1}
\begin{align}
 & 2\phi_{2}^{(n)}\phi_{2}-(\phi_{2}^{(n)})^{2}\geq\mathrm{Tr}\left(\mathbf{\Theta}\right)\left(1-z\right),\label{eq:27-1}\\
 & \left[\begin{array}{cc}
z & \phi_{2}\\
\phi_{2} & \mathrm{Tr}(\stackrel[i=1]{2}{\sum}\mathbf{W}_{e}^{i})
\end{array}\right]\succeq\mathbf{0},\label{eq:28}
\end{align}

\end{subequations}

\noindent where $z=e^{\tau^{(n)}}(\tau-\tau^{(n)}+1)$. At this point,
constraint (\ref{eq:21}) is replaced by two convex constraints, i.e.,
the convex constraint (\ref{eq:27-1}) and the linear matrix inequality
constraint (\ref{eq:28}).

So far, the optimization problem (\ref{max_1-1}) has been reformulated as
\begin{subequations}\label{max_3}
\noindent
\begin{align}
\underset{\begin{array}{c}
\varphi_{2},\begin{array}{c}
\tau,\mu_{i},\end{array}\\
\beta,\xi,\phi_{2}
\end{array}}{\mathrm{max}} & \stackrel[i=1]{2}{\sum}\mu_{i}-\tau\label{eq:25a}\\
 \mathrm{s.t.\quad\quad\:} & (\ref{eq:13}),(\ref{eq:15-1}),(\ref{eq:15-1-1}),(\ref{FOTE}),\label{eq:25b}\\
 & (\ref{FOTE1}),(\ref{eq:26-1}),\,\mathrm{and}\,(\ref{FOTE1-1}),\label{eq:25c}\\
 & \mathbf{V}_{i}\succeq\mathbf{0},\mu_{i}\geq0,i\in\left\{ 1,2\right\} .\label{eq:25f}
\end{align}

\end{subequations}

\noindent  Consequently, for the  active Eves case, the original optimization problem (\ref{max_1}) has been
transformed into the convex form. Then the solution $(\mathbf{V}_{1}^{*},\mathbf{V}_{2}^{*},\mathbf{W}^{*})$
of (\ref{max_1}) is obtained by off-the-shelf optimization solver, e.g., SeDuMi
and Yalmip. The algorithmic implementation of the proposed OSBD scheme is summarized in Algorithm 2. Note that problem
(\ref{max_3}) is handled iteratively with the initial value of $(\tau^{(0)},\mu_{1}^{(0)},\mu_{2}^{(0)},\varphi_{2}^{(0)},\phi_{2}^{(0)},\xi^{(0)}).$
For each iteration, $(\tau^{(n)},\mu_{1}^{(n)},\mu_{2}^{(n)},\varphi_{2}^{(n)},\phi_{2}^{(n)},\xi^{(n)})$
is updated with the solution obtained in the previous  iteration until
the rate gap between two successive iterations below the predefined
accuracy. Moreover, the optimal solution $(\mathbf{V}_{1}^{*},\mathbf{V}_{2}^{*},\mathbf{W}^{*})$
yielded by SDR is rank-one, which is described in the following proposition.

\begin{algorithm}[t]
\caption{OSBD Scheme with Eves' CSI}

1: \textbf{Input:} Set $n=0$, $\tau^{(0)}=1$, $\mu_{1}^{(0)}=1,\,\mu_{2}^{(0)}=0,$

$\quad\,\varphi_{2}^{(0)}=1,\,\phi_{2}^{(0)}=1,$ $\xi^{(0)}=1,$
$\epsilon=10^{-4}$, and

$\quad\,$$R_{0}=\mu_{1}^{(0)}+\mu_{2}^{(0)}-\tau^{(0)}.$

2: \textbf{Repeat }

a) Solve (\ref{max_3}) and obtain $\varphi_{2}^{*},\phi_{2}^{*},\xi^{*}$
as well as

$\quad\:$the SSR, i.e., $R=\mu_{1}^{*}+\mu_{2}^{*}-\tau^{*}.$

b) Update $\eta^{(n)}$ based on (\ref{eq:23}).\\
c) Update $(\tau^{(n)},\mu_{1}^{(n)},\mu_{2}^{(n)},\varphi_{2}^{(n)},\phi_{2}^{(n)},\xi^{(n)})$$\leftarrow$

$\quad\,$$(\tau^{*},\mu_{1}^{*},\mu_{2}^{*},\varphi_{2}^{*},\phi_{2}^{*},\xi^{*})$,
$R_{0}\leftarrow R$.

d) Set $n\leftarrow n+1$.

3: \textbf{Until} $\left|R-R_{0}\right|^{2}\leq\epsilon$.

4: \textbf{Output: $\beta^{*},\mathbf{V}_{1}^{*},$ $\mathbf{V}_{2}^{*}$
}and $\mathbf{W}$.
\end{algorithm}

\begin{prop}
If problem (\ref{max_3}) is feasible, the beamforming vectors $\mathbf{v}_{1}$
, $\mathbf{v}_{2}$ and $\mathbf{w}$ can be exactly obtained by eigenvalue
decomposition of $\mathbf{V}_{1}^{*}, \mathbf{V}_{2}^{*}$ and $\mathbf{W}^{*}$, since $\mathbf{V}_{1}^{*},\mathbf{V}_{2}^{*}$ and $\mathbf{W}^{*}$
are rank-one.
\end{prop}
\begin{IEEEproof}
The transmit beamforming vectors are jointly optimized through the OSBD scheme. According to the rank reduction procedure of the semidefinite programming given in Lemma 3.1 of \cite{rank_one}, $\mathbf{V}_{1}^{*}$ , $\mathbf{V}_{2}^{*}$ and
$\mathbf{W}^{*}$ satisfy the following inequality

\begin{align}
 & \mathrm{rank}^{2}(\mathbf{V}_{1}^{*})+\mathrm{rank}^{2}(\mathbf{V}_{2}^{*})+\mathrm{rank}^{2}(\mathbf{W}^{*})\leq3.
\end{align}

\noindent Since from (\ref{eq:17d-1-2}) and (\ref{eq:25b-2}) it is observed
that problem (\ref{max_3}) is not feasible if $\mathbf{V}_{1}^{*}=\mathbf{0}$
or $\mathbf{V}_{2}^{*}=\mathbf{0}$, then, $\mathrm{rank}(\mathbf{V}_{1}^{*})=\mathrm{rank}(\mathbf{V}_{2}^{*})=1,$
and also $\mathrm{rank}(\mathbf{W}^{*})\leq1$.  Since $\alpha>0$,
$\mathrm{rank}(\mathbf{W}^{*})\neq0,$ then clearly there exits a $\mathbf{W}^{*}$ which satisfies $\mathrm{rank}(\mathbf{W}^{*})=1$ and concludes the proof.
\end{IEEEproof}

\subsection*{B. Single-Antenna Configuration}

In applications where devices are equipped with a single antenna, the secure beamforming design problem analyzed in the previous subsection, simplifies to a power allocation problem. As there will be no diversity gain and in order to maximize the SSR and at the same time avoiding the AN interference to $D_2$, all the energy harvested by $D_1$ is used for information transmission. In other words, since the transmitted signal from $D_{1}$ during the second phase is represented as $x=\sqrt{P_{t}}s_{2}$, the EH constraint
from (\ref{max_1}) can be removed. Then, under the constraints of
the transmit power at $S$ and the QoS requirement at $D_{i}$, the SSR is optimized. By conveniently introducing power allocation factors $\rho_{1}$ and $\rho_{2}$ as replacement of the beamforming vectors $\mathbf{v}_{1}$ and $\mathbf{v}_{2}$,
respectively, the optimization problem is reformulated as

\begin{subequations}\label{max_1-1-1}

\noindent
\begin{align}
\underset{\beta,\rho_{1},\rho_{2}}{\mathrm{max}}\, & \mathrm{log}\left(1+P_{s}\left(1-\beta\right)h_{1}\right)-\nonumber\\
 & \mathrm{log}\left(1+\frac{\left\Vert \mathbf{g}_{se}\right\Vert ^{2}P_{s}+P_{s}\beta h_{1}\left\Vert \mathbf{g}_{1e}\right\Vert ^{2}}{2\sigma^{2}}\right)\label{eq:17a-1-1}\\
\mathrm{s.t.\;\:\,} & P_{s}\left(1-\beta\right)h_{1}\rho_{1}\geq\gamma_{1},\label{eq:17b-1-1}\\
 & \frac{PH_{s}\left(1-\beta\right)h_{1}\rho_{2}}{1+P_{s}\left(1-\beta\right)h_{1}\rho_{1}}\geq\gamma_{2},\label{eq:17c-1-1}\\
 & \frac{P_{s}h_{2}\rho_{2}}{1+P_{s}h_{2}\rho_{1}}+P_{s}\beta h_{12}h_{1}\geq\gamma_{2},\label{eq:17d-1-1}\\
 & \rho_{1}+\rho_{2}=1,\rho_{1},\rho_{2}\in[0,1],
\end{align}

\end{subequations}

\noindent where $h_{1},h_{2}$ and $h_{12}$ are normalized main channel
gains; $\mathbf{g}_{se}$ and $\mathbf{g}_{1e}$ are wiretap channels;
$\gamma_{i},$ is target SINR of $D_{i}$; (\ref{eq:17b-1-1}) and (\ref{eq:17d-1-1}) are QoS constraints of $D_{i}$; And (\ref{eq:17c-1-1})
ensuring that $D_1$ can satisfactorily perform SIC. As this optimization problem (\ref{max_1-1-1}) is
a bilevel programming problem, with the outer level variable being
$\beta$, it will be transformed into the convex one.

Following \cite{SCA}, it can be shown that at least
one inequality constraint in (\ref{max_1-1-1}) is satisfied with
equality at the optimal solution when problem (\ref{max_1-1-1}) is
feasible. Otherwise, the optimization variables can be changed to
increase the SSR until one of the constraints holds with equality.

For convenience of the analysis' presentation, let us consider that the inequality
constraint (\ref{eq:17b-1-1}) holds with equality at the optimal
solution so that
\begin{align}
 & \rho_{1}=\frac{\gamma_{1}}{P_{s}\left(1-\beta\right)h_{1}},\rho_{2}=\frac{P_{s}\left(1-\beta\right)h_{1}-\gamma_{1}}{P_{s}\left(1-\beta\right)h_{1}}.\label{eq:23a-1-1-1}
\end{align}

\noindent It is noted that the variables $\rho_{1}$ and $\rho_{2}$ are
expressed as a function of $\beta$. Substituting (\ref{eq:23a-1-1-1})
into (\ref{eq:17c-1-1}) and (\ref{eq:17d-1-1}), constraints (\ref{eq:17c-1-1})
and (\ref{eq:17d-1-1}) are rewritten as

\begin{subequations}\label{eq:constraint}

\begin{align}
 & P_{s}\left(1-\beta\right)h_{1}\geq\gamma_{1}+\gamma_{2}+\gamma_{1}\gamma_{2},\label{eq:equality}\\
 & P_{s}h_{1}^{2}h_{12}\beta^{2}\leq\xi+\zeta,\label{eq:equality1}
\end{align}

\end{subequations}

\noindent respectively, where $\xi=P_{s}h_{1}h_{2}-\gamma_{2}h_{1}-\gamma_{1}\gamma_{2}h_{2},\zeta=\beta\left(\gamma_{2}h_{1}-P_{s}h_{1}h_{2}+P_{s}h_{1}h_{2}h_{12}\gamma_{1}+P_{s}h_{1}^{2}h_{12}\right).$
Since the optimization, problem (\ref{max_1-1-1}) due to is still non-convex, a slack variable u is introduced to non-convex objective approximate the second term in (\ref{eq:17a-1-1})
by FOTE. In this way, the non-convex optimization problem (\ref{max_1-1-1})
can be reformulated as

\begin{subequations}\label{max_SISO}

\noindent
\begin{align}
\underset{\beta,\upsilon}{\mathrm{max}}\, & \,\,\mathrm{log}\left(1+P_{s}\left(1-\beta\right)h_{1}\right)-\upsilon\label{eq:17a-1-1-1}\\
\mathrm{s.t.\;\,} & 1+\frac{\left\Vert \mathbf{g}_{se}\right\Vert ^{2}P_{s}+P_{s}\beta h_{1}\left\Vert \mathbf{g}_{1e}\right\Vert ^{2}}{2\sigma^{2}}\leq z,\label{eq:17b-1-1-1}\\
 & (\ref{eq:equality})\,\mathrm{and}\,(\ref{eq:equality1}),
\end{align}

\end{subequations}
\noindent where $z=e^{\upsilon^{(n)}}\left(\upsilon-\upsilon^{(n)}+1\right).$
Since the optimization problem stated in (\ref{max_SISO}) is now in convex
form, its optimal solution can be readily obtained by the handy solver, e.g., SeDuMi
or Yalmip. Note that if (\ref{eq:17c-1-1}) or (\ref{eq:17d-1-1}) hold with equality, a similar methodology can be followed to obtain an equivalent convex optimization problem. However, due to space limitations, the detailed procedure will not be presented here.

\section{Performance evaluation results and discussion }

In this section, the SSR performance of the previously described cooperative relay-aided secure IoT communication systems will be presented. The various performance evaluation results have been obtained by extensive computer simulation experiments.
It is assumed that the two
legitimate users and $N_{e}$ eavesdroppers are deployed within a 8-meter$\times$8-meter
network, while $S$ is located at the edge with coordinate $(0,4)$.
$h_{Li}=10^{-3}d_{i}^{-\alpha_{i}},(i=1,2)$ and $h_{Le}=10^{-3}d_{e}^{-\alpha_{e}}$
represent the large-scale path losses, where $d_{i},(i=1,2)$ and $d_{e}$
are the distances from $S$ to $D_{i}$ and Eves, respectively, $\alpha_{1}=2,\alpha_{2}=4$
and $\alpha_{e}=2$ are the corresponding path loss exponents. The
downlink channels $\mathbf{H}_{s1},\mathbf{H}_{12},\mathbf{G}_{se}$
and $\mathbf{G}_{1e}$ operate in the presence of Rician fading, as follows

\noindent
\begin{align}
 & \mathbf{H}=\sqrt{\frac{\Gamma}{1+\Gamma}}\mathbf{H}^{LoS}+\sqrt{\frac{1}{1+\Gamma}}\mathbf{H}^{NLoS},\label{eq:}
\end{align}

\noindent where $\mathbf{H}^{LoS}$ is the LOS deterministic component,
$\mathbf{H}^{NLoS}$ is modeled as the small-scale Rayleigh fading
following $\mathcal{CN}(0,1)$, and  unless otherwise stated, with Rician factor $\Gamma=4$. The distribution of $\mathbf{H}_{s2}$ follows $\mathcal{CN}(0,1)$.
Matlab is used as a simulation tool, and Yalmip is employed as an optimization  solver.
All the SSR curves are generated by averaging 500 independent channel realizations.

\subsection*{A. Performance Evaluation of Proposed AN-aided Scheme Without Eves' CSI  }

Fig. 2  presents  the SSR versus the number of iterations for different $N_{1}$  and $r_{1}$, where the SINR threshold $\gamma={2,3}$ in Fig. 2(a) and  Fig. 2(b), respectively. We set the number of eavesdroppers, $N_{e}=3$, the transmit power constraint at $S$,$P_{s}=30\ \mathrm{dBm}$.
These performance evaluation results clearly show  that the proposed ASBD scheme converges to a stationary point in few iterations.  Furthermore,  the SSR performance improves as  $\gamma$  (i.e., the QoS requirement of $D_{2}$) decreases. This happens because   the harvested power at $D_{1}$ is larger if $\gamma$ become smaller, and  then more power can be used to transmit AN.

Fig. 3 presents the SSR performance of  the ASBD scheme versus the transmit power at $S$ for different values of $N_{1}$  by fixing $N_{s}=N_{2}=4$, $N_{e}=3$ and $\gamma=3$.
For comparison purposes, the performance of the secure beamforming design scheme without AN-aided (denoted as ``Without AN'')  has been also obtained.
These results clearly show  that the SSR obtained by ASBD scheme is larger compared with the benchmarks. Furthermore,
it is observed that, with the increase of $N_{1}$, the SSR achieved by different schemes is improved significantly.

Fig. 4 shows the SSR  versus the predefined threshold of achievable ergodic rate of $D_{1}$ for different values of $N_{s}$, where $P_{s}=30\ \mathrm{dBm}$, $N_{1}=N_{2}=4$, $\gamma=3$. It can be seen  that the SSR performance is enhanced as $r_{1}$ increases, which is consistent with previous findings presented  in \cite{fairness}. The reason is that  the transmitter $S$ allocates more power to the strong user $D_{1}$ to satisfy the predefined threshold of achievable ergodic rate of $D_{1}$, which leads to the increasing of SSR.  Meanwhile, compared with the OMA-based secure beamforming design with AN-aided
(denoted as ``OMA w/ AN''),  the proposed design exhibits  better SSR performance when $N_{s}$ increases.

\begin{figure}[t]
\centering{}\subfloat[]{\includegraphics[scale=0.33]{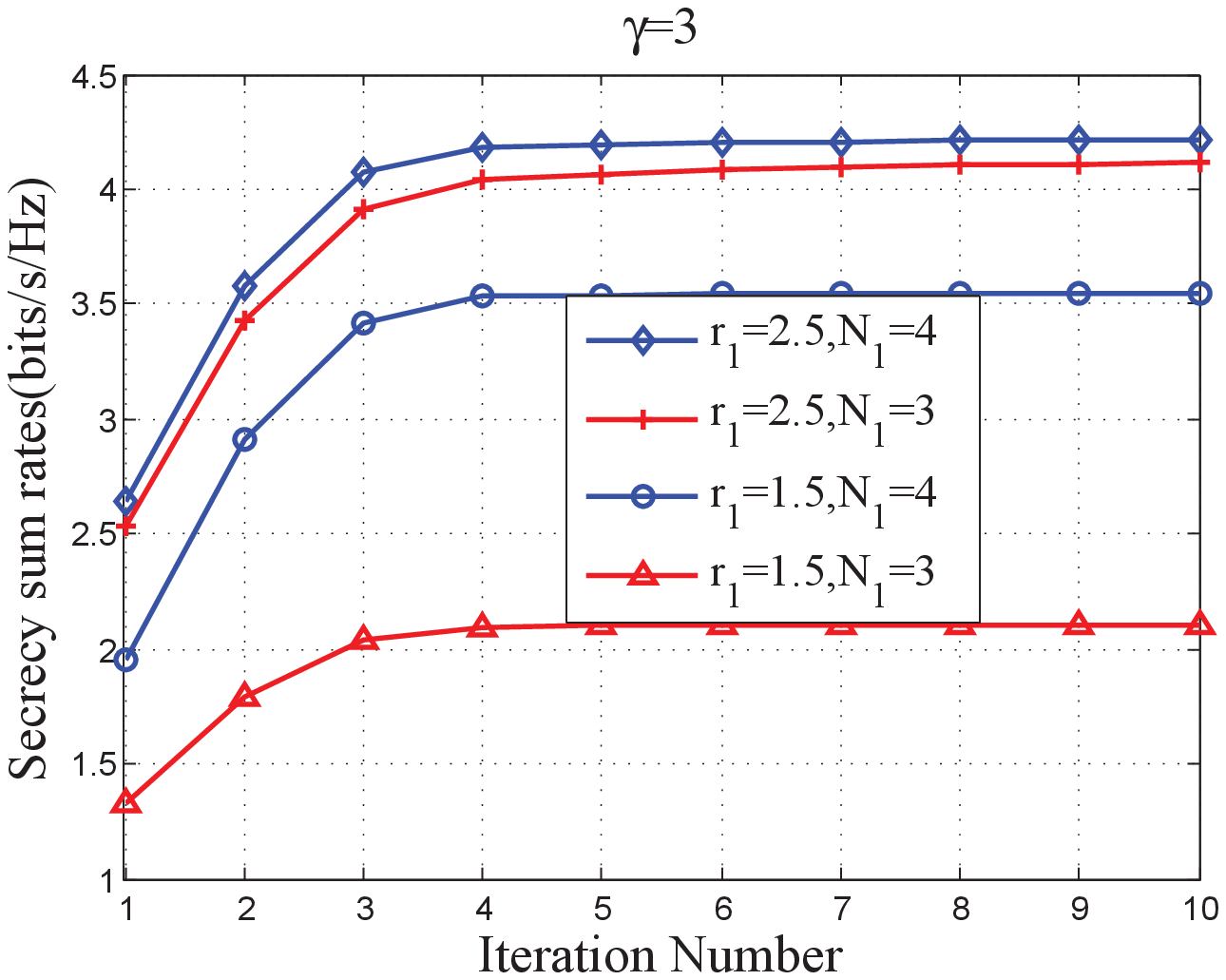}

}\subfloat[]{\includegraphics[scale=0.32]{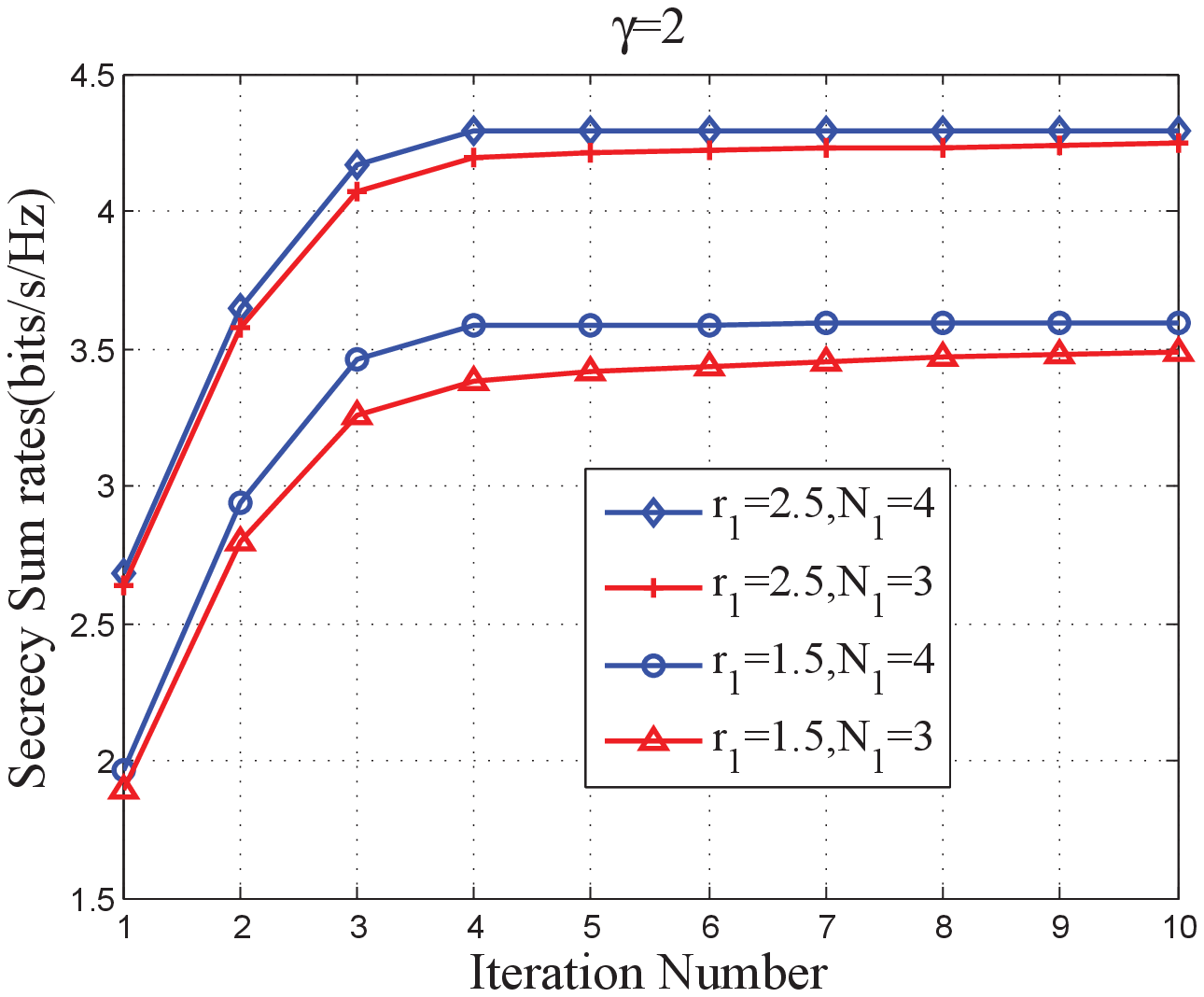}
}

\caption{SSR versus the number of iteration for different $N_{1}$s and $r_{1}$s. (a) $\gamma=3$. (b) $\gamma=2$.}
\end{figure}

\begin{figure}[t]
\centering{}

\includegraphics[scale=0.63]{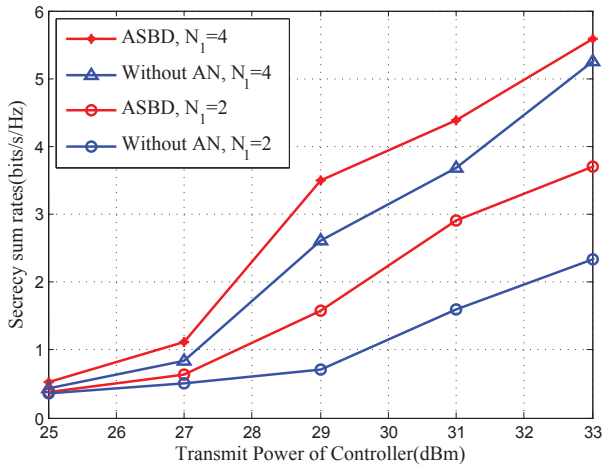}

\caption{SSR versus $P_{s},$ comparisons of the  proposed ASBD scheme and the scheme without AN-aided, for different $N_{1}$s.}
\end{figure}

\begin{figure}[t]
\centering{}

\includegraphics[scale=0.63]{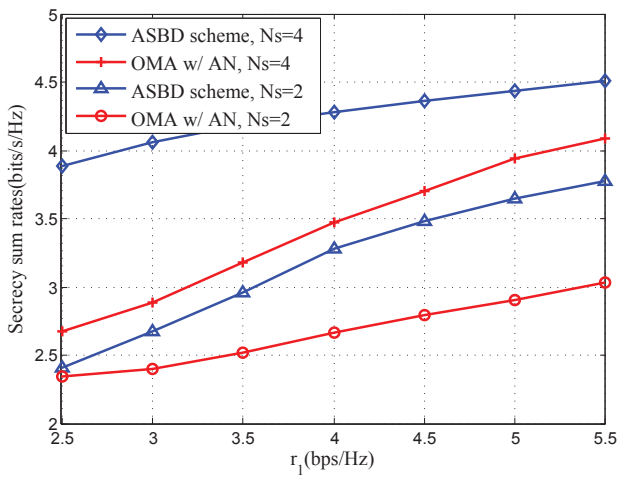}

\caption{SSR versus $r_{1},$ for different $N_{s}$.}
\end{figure}

\subsection*{B. SSR Comparison With  Eves' CSI}

In Fig. 5, the convergence property of the proposed
OSBD algorithm and the algorithm given in \cite{Secrecy_sum_rate}
for the active Eves case is presented,  with  $P_{s}=30\ \mathrm{dBm},$ $\gamma=3$,  $N_{e}=3,N_{i}=4,N_{s}=\{2,4\}.$
It is noted that, although the algorithm of \cite{Secrecy_sum_rate}  also employs a NOMA transmission protocol and optimizes the SSR of downlink systems,
it only considers the direct transmission link and does not use any physical layer security technology.
From Fig. 5, consensus has been reached that the proposed algorithm
converges to a stationary point in a fewer steps than that in \cite{Secrecy_sum_rate},
especially for higher values of $N_{s}$. To demonstrate the computational efficiency of the proposed approach, we give the time needed for the two algorithms to converge. Our proposed algorithm takes about 4.75 s for each independent channel realizations, while the time required for the algorithm given in [2] is about 7.11 s.

Fig. 6 illustrates the SSR performance of the proposed OSBD
scheme for the active Eves case. Set $N_{e}=3,N_{i}=4,N_{s}=\{2,4\},\gamma=3$.
We also consider the secure transmission design given in \cite{Secrecy_sum_rate},
the NOMA secure transmission protocol without relay-aided (denoted as
``NOMA w/o EH'') as well as the conventional OMA-based secure beamforming design without relay-aided (denoted
as ``OMA w/o EH'') as  baseline algorithms. As expected, the proposed OSBD scheme based on the cooperative secure transmission protocol can obtain higher secrecy
rate than that in \cite{Secrecy_sum_rate}. In addition, numerical results also indicate
that the proposed scheme can achieve much better SSR performance than
the other two benchmark schemes (i.e., NOMA w/o EH scheme and OMA w/o EH scheme). With the fixed $P_{s}$, we can observe that, as $N_{s}$ grows, the SSR performance becomes better in different
schemes, since more array gains are provided.

Fig. 7 illustrates the SSR performance under different transmit powers of controller for the single-antenna configuration. Similar to Fig. 6, the three benchmarks are employed as baseline algorithms, where $N_{1}=N_{2}=N_{s}=1, N_{e}=3, \gamma_{1}=2,\gamma_{2}=3$. Obviously, the SSR performance of the proposed power control scheme outperforms that of the other three benchmarks, especially in large transmit power region. The reason is that the received signal at $D_{2}$
is consisted of two parts from the relay-assisted system so that the SSR of the IoT is further enhanced.

\begin{figure}[t]
\centering{}

\includegraphics[scale=0.63]{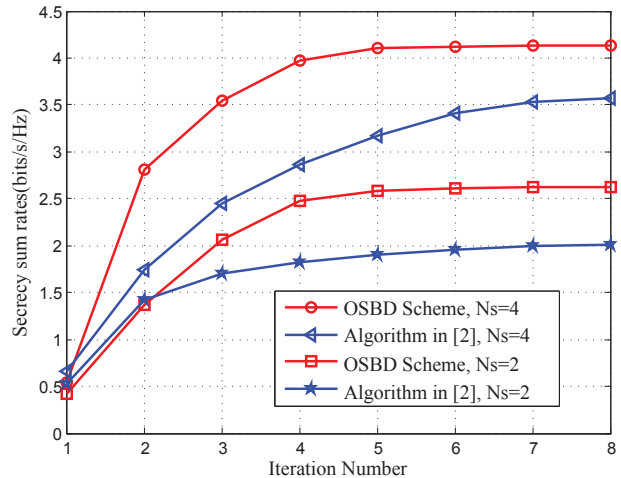}

\caption{The SSR performance versus the number of iteration for different $N_{s}$s. }
\end{figure}

\begin{figure}[t]
\centering{}

\includegraphics[scale=0.63]{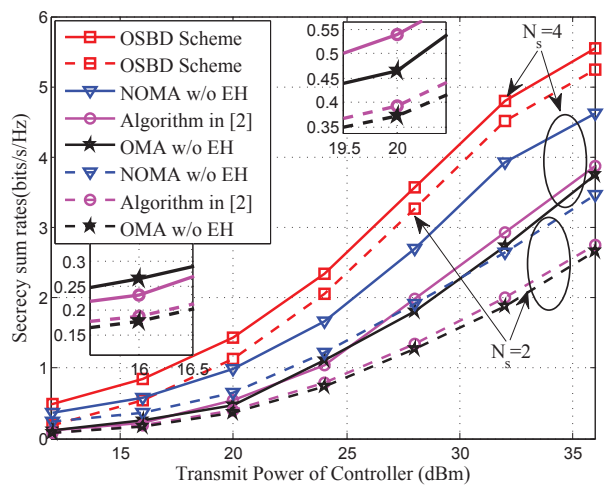}

\caption{The SSR performance of different schemes versus $P_{s}$.}
\end{figure}

\begin{figure}[t]
\centering{}

\includegraphics[scale=0.63]{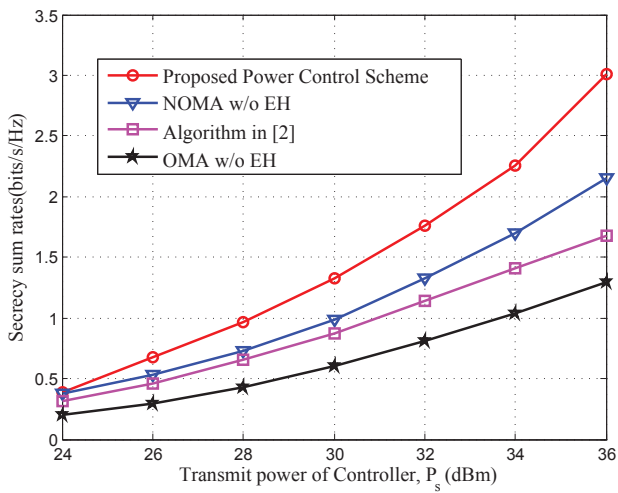}

\caption{The SSR performance of different schemes versus $P_{s}$  with single-antenna configurations.}
\end{figure}

\section{Conclusions}

In this paper, we investigated the secure transmit beamforming design in the downlink
IoT system. Taking diversified QoS requirements of IoT users into
account, a  novel  relay-aided cooperative  secure transmission strategy was proposed
to further improve the SSR of the IoT system. Two Eves cases have
been considered based on the availability of the  CSIs of Eves. For the passive Eves case,  ASBD scheme was proposed to jointly optimize the secure
beamforming vectors, AN covariance matrix as well as PS ratio. For the active multi-antennas Eves case,
OSBD scheme was proposed to maximize the SSR of the considered IoT system. When the transmitters (i.e., $S$ and $D_{1}$) are single-antenna configurations, the orthogonal-projection-based optimization problem is degraded into a power allocation problem and the optimal solution can be found by fully exploiting the specific property of the optimization problem. Various performance evaluation results have  demonstrated the superiority of the proposed schemes compared to the existing benchmarks.

The proposed cooperative SWIPT secure transmission protocol can be applied to massively connected IoT systems in the future, such as Healthcare IoT, in which a large quantity of smart devices are involved in sensing health parameters and these devices are confined by the limited power. For Healthcare IoT applications, it is noted that  security is a vital yet challenging requirement  during data collection from patients to a centralized data collection center. Thus, as  the proposed secure transmission scheme is capable of ensuring the security of Healthcare IoT, it could be very well  used for such kind of applications.

\bibliographystyle{IEEEtran}

\bibliography{ciations}
\begin{IEEEbiography}[{\includegraphics[width=1in,height=1.25in,clip,keepaspectratio]{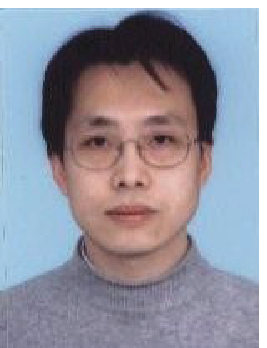}}]{Pingmu Huang}

received the M.S. degrees from Xi¡¯an Jiaotong University, Xian, China, in 1996 and received Ph.D. degree of Signal and Information Processing from Beijing University of Posts and Telecommunications (BUPT), Beijing, China, in 2009. He is now a lecturer with the School of Information and Communication Engineering, BUPT. His current research interests include speech synthesis and signal processing. He published several journal papers and conference papers on speech synthesis and signal processing.

\end{IEEEbiography}
\begin{IEEEbiography}[{\includegraphics[width=1in,height=1.25in,clip,keepaspectratio]{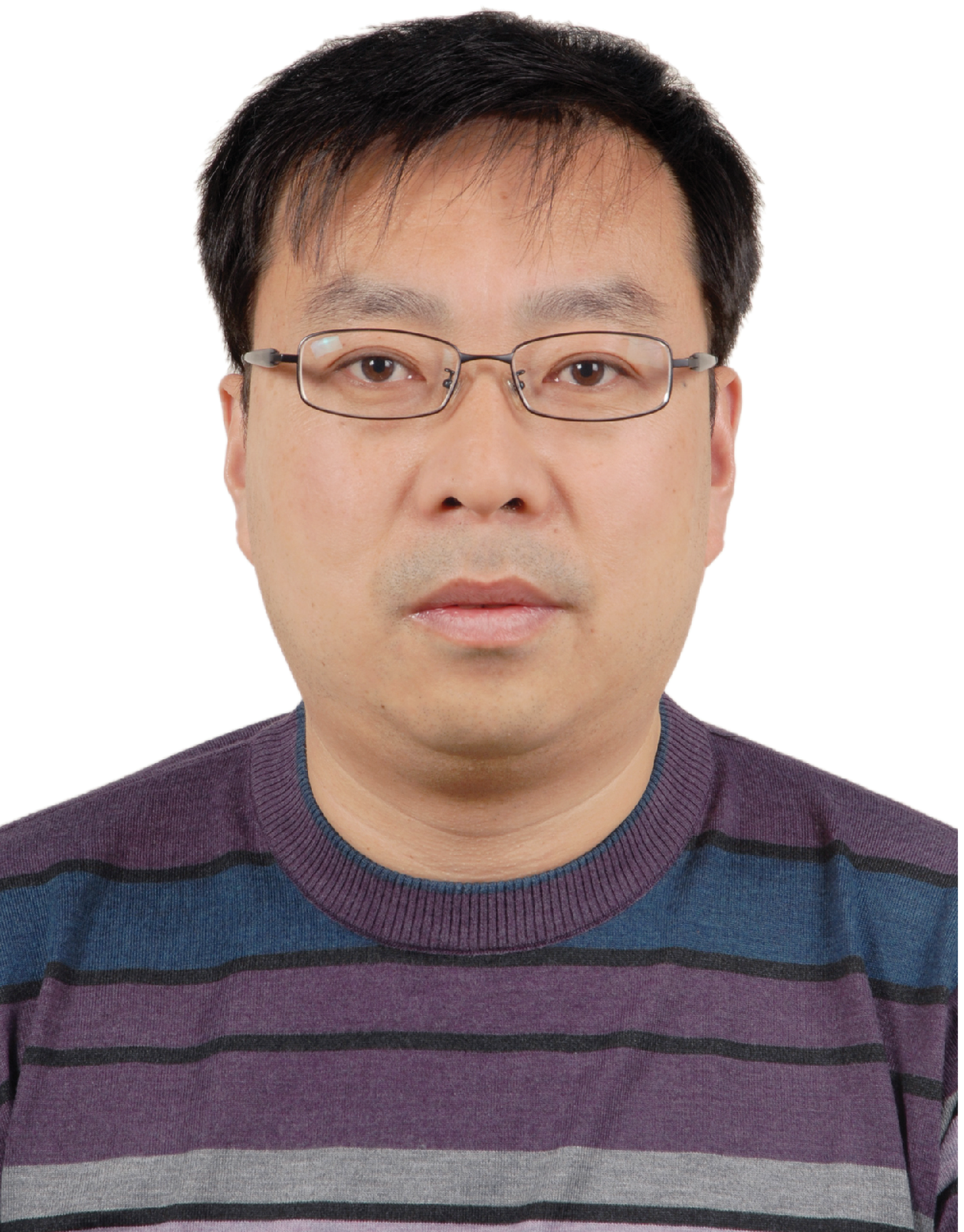}}]{Tiejun Lv}

(M'08-SM'12) received the M.S. and Ph.D. degrees in electronic engineering from
the University of Electronic Science and Technology of China (UESTC), Chengdu, China, in 1997 and 2000, respectively.
From January 2001 to January 2003, he was a Postdoctoral Fellow with Tsinghua University, Beijing, China.
In 2005, he was promoted to a Full Professor with the School of Information and Communication Engineering,
Beijing University of Posts and Telecommunications (BUPT).
From September 2008 to March 2009, he was a Visiting Professor with the Department of Electrical Engineering, Stanford University, Stanford, CA, USA.
He is the author of 2 books, more than 60 published IEEE journal papers and 180 conference papers on the physical layer of wireless mobile communications.
His current research interests include signal processing, communications theory and networking.
He was the recipient of the Program for New Century Excellent Talents in University Award from the Ministry of Education, China, in 2006.
He received the Nature Science Award in the Ministry of Education of China for the hierarchical cooperative communication theory and technologies in 2015.

\end{IEEEbiography}

\end{document}